\documentclass[a4paper,11pt]{article}

\usepackage[utf8]{inputenc}
\usepackage{geometry}
\usepackage{graphicx}
\usepackage{color}
\usepackage{amsmath}
\usepackage{amssymb}
\usepackage{lineno}
\usepackage{enumitem}
\usepackage{array}
\usepackage{upgreek}
\usepackage{siunitx}
\usepackage[normalem]{ulem}
\usepackage{subcaption}
\usepackage{float}
\usepackage{titlesec}
\usepackage{booktabs}
\usepackage[style=numeric,sorting=none]{biblatex}
\usepackage{hyperref}
\usepackage{setspace}
\usepackage{multirow}
\usepackage{rotating}
\usepackage{placeins}
\usepackage{authblk}
\usepackage{threeparttable}

\def\p{\partial}
\def\d{\mathrm{d}}
\def\abs#1{\left\vert{#1}\right\vert}
\def\order#1{\mathcal{O}({#1})}

\def\ybar{\overline{y}}
\def\Reynolds{\text{\textit{Re}}}
\def\Darcy{\text{\textit{Da}}}
\def\Prandtl{\text{\textit{Pr}}}
\def\Mach{\text{\textit{Ma}}}

\def\averageinterphasefluxfs#1{\left\langle#1\right\rangle_{\partial fs}}
\def\fluid#1{\left\langle#1\right\rangle_f}
\def\solid#1{\left\langle#1\right\rangle_s}

\DeclareRobustCommand{\VAN}[3]{#2}

\title{\textbf{Compressible boundary layers \\over isotropic porous surfaces}}

\author[1]{Ludovico Foss\`{a}\footnote{Corresponding author ludovico-fossa@oist.jp. Present address: Complex Fluids and Flows Unit, Okinawa Institute of Science and Technology Graduate University, 1919-1 Tancha, Onna-son, Kunigami-gun, Okinawa-ken 904-0495, Japan.}}
\author[1]{Pierre Ricco}

\affil[1]{School of Mechanical, Aerospace and Civil Engineering, Mappin Building, University of Sheffield, S13JD Sheffield, United Kingdom}

\newcommand{\bluealert}[1]{\textcolor{blue}{\textbf{#1}}}
 
\DeclareRobustCommand{\VAN}[3]{#2}

\begin{document}

\DeclareRobustCommand{\VAN}[3]{#3}

\maketitle
\onehalfspacing

\bluealert{Accepted for publication on Physical Review Fluids}

\begin{abstract}
A compressible laminar boundary layer developing over an isotropic porous substrate is investigated by asymptotic and numerical methods. The substrate is modeled as an array of cubes. The momentum and enthalpy balance equations are derived by volume averaging. The self-similar solution proposed by Tsiberkin (2018) [\emph{Transp. Porous Media} 121(1):109–120] for streamwise-growing permeability is extended to include compressibility, heat conduction and a nonlinear drag. The velocity profile shows an inflection point at the free fluid-porous interfacial layer, below which it decreases to zero. A marked reduction of the adiabatic recovery temperature of the fluid and the velocity gradient at the interface is observed for high porosity, large grains and relatively high Mach numbers. The temperature imposed at the bottom of the porous substrate has a negligible influence on the shear stresses.
\end{abstract}

\section{Introduction}

Fluid flows over saturated porous media are common in industrial and engineering contexts \cite{Nield_Bejan_2017,Vafai_Kim_1990,Nield_Kutsenov_2003,Breugem_Boersma_Uittenbogaard_2005,Tsiberkin_2016}. Several theoretical, numerical and experimental studies have focused on the coupling between a free fluid and an adjacent saturated porous medium in confined geometries \cite{Breugem_Boersma_Uittenbogaard_2006,Wu_Mirbod_2018,Harter_Martinez_Poser_Weigand_Lamanna_2023} and on wall-bounded flows through a semi-infinite porous medium \cite{Kaviany_1987,Nakayama_Kokudai_Koyama_1990,Papalexandris_2023a}. Much less attention has been devoted to the coupling between an unbounded fluid and a porous medium. \cite{Vafai_Kim_1990} studied the effects of a porous substrate on the heat transfer and the drag exerted by an overflowing fluid subject to a streamwise pressure gradient. They solved the steady, two-dimensional Navier-Stokes equations with a Darcy term and a Forchheimer term. The boundary-layer approximation was not imposed and the governing equations were elliptic, yet their results featured a marked parabolic character. \cite{Nield_Kutsenov_2003} modeled the influence of a uniform porous substrate on an incompressible Blasius boundary layer by expanding the streamfunction in a power series of $(K^\ast/x^\ast)^{1/2}$, where $x^\ast$ is the streamwise coordinate and $K^\ast$ is a sufficiently small permeability. They implicitly assumed continuity in the velocity and shear stresses across the fluid-porous interface \cite{Neale_Nader_1974}. Using a similar approach, \cite{Breugem_Boersma_Uittenbogaard_2005} proposed a theoretical model for an incompressible laminar boundary layer over a porous flat plate and adopted the interfacial conditions of \cite{Ochoa-Tapia_Whitaker_1995a}. The pores were assumed to be small enough for the non-Darcian effects to be negligible below the interface. 

Theoretical and numerical analyses of compressible flows within porous media have drawn much less attention. \cite{Nield_1994} used the volume-averaging approach \cite{Bear_Bachmat_1990,Whitaker_1998} to study the effect of choking and included the Darcy, Forchheimer and inertial terms in his analysis. The different roles of the advection and Forchheimer terms in compressible flows were also discussed by \cite{de-Ville_1996}. Porous substrates have been used to delay boundary-layer transition to turbulence in supersonic wind tunnels \cite{Mironov_Maslov_Poplavskaya_Kirilovskiy_2015,Maslov_Mironov_Poplavskaya_Kirilovskiy_2019,Running_Bemis_Hill_Borg_Redmond_Jantze_Scalo_2023}. However, most studies have focused on the attenuation of small-amplitude acoustic disturbances in flat-plate boundary layers rather than the modification of the velocity profiles of the laminar base flows \cite{Maslov_Mironov_Poplavskaya_Kirilovskiy_2019}. 

The modelling and computation of boundary-layer flows over and within porous surfaces of constant thickness and uniform permeability pose significant challenges, primarily because the thickness of the boundary layer grows downstream. Self-similar solutions cannot be derived when uniform porous substrates are considered \cite{Vafai_Kim_1990,Papalexandris_2023a}. The streamwise-marching computations have relied on the existence of a self-similar solution upstream \cite{Sparrow_Quack_Boerner_1970,Cebeci_2002}. 

A similarity solution was found by \cite{Tsiberkin_2016,Tsiberkin_2018a,Tsiberkin_2018b} for the case of an incompressible boundary layer on a porous substrate of infinite thickness and streamwise-increasing permeability. His analysis did not take the nonlinear Forchheimer term into account. He concluded that neither the Brinkman nor the advection term can be neglected without losing important features of the momentum transfer and interfacial stability. Albeit stemming from an idealized setting, self-similar and local-similarity solutions permit to overcome all these issues and to unravel physical mechanisms that are relevant to more realistic scenarios, particularly when the effects of compressibility are important \cite{Stewartson_1964,Anderson_2019}. The central aim of the present work is to extend Tsiberkin's solution to high-speed boundary layers. For the first time, the governing equations are derived by volume averaging and the effects of nonlinear drag, compressibility and heat conduction are found numerically. 

\FloatBarrier

%%%%%%%%%%%%%%%%%%%%%%%%%%%%%%%%%%%%%%%%%%%%%%%%%%%%
%%%%%%%%%%%%%%%%%%%%%%%%%%%%%%%%%%%%%%%%%%%%%%%%%%%%
%%%%%%%%%%%%%%%%%%%%%%%%%%%%%%%%%%%%%%%%%%%%%%%%%%%%
%%%%%%%%%%%%%%%%%%%%%%%%%%%%%%%%%%%%%%%%%%%%%%%%%%%%
%%%%%%%%%%%%%%%%%%%%%%%%%%%%%%%%%%%%%%%%%%%%%%%%%%%%

\section{Mathemetical framework}

\begin{figure}[ht]
    \centering
    \includegraphics[width=\linewidth]{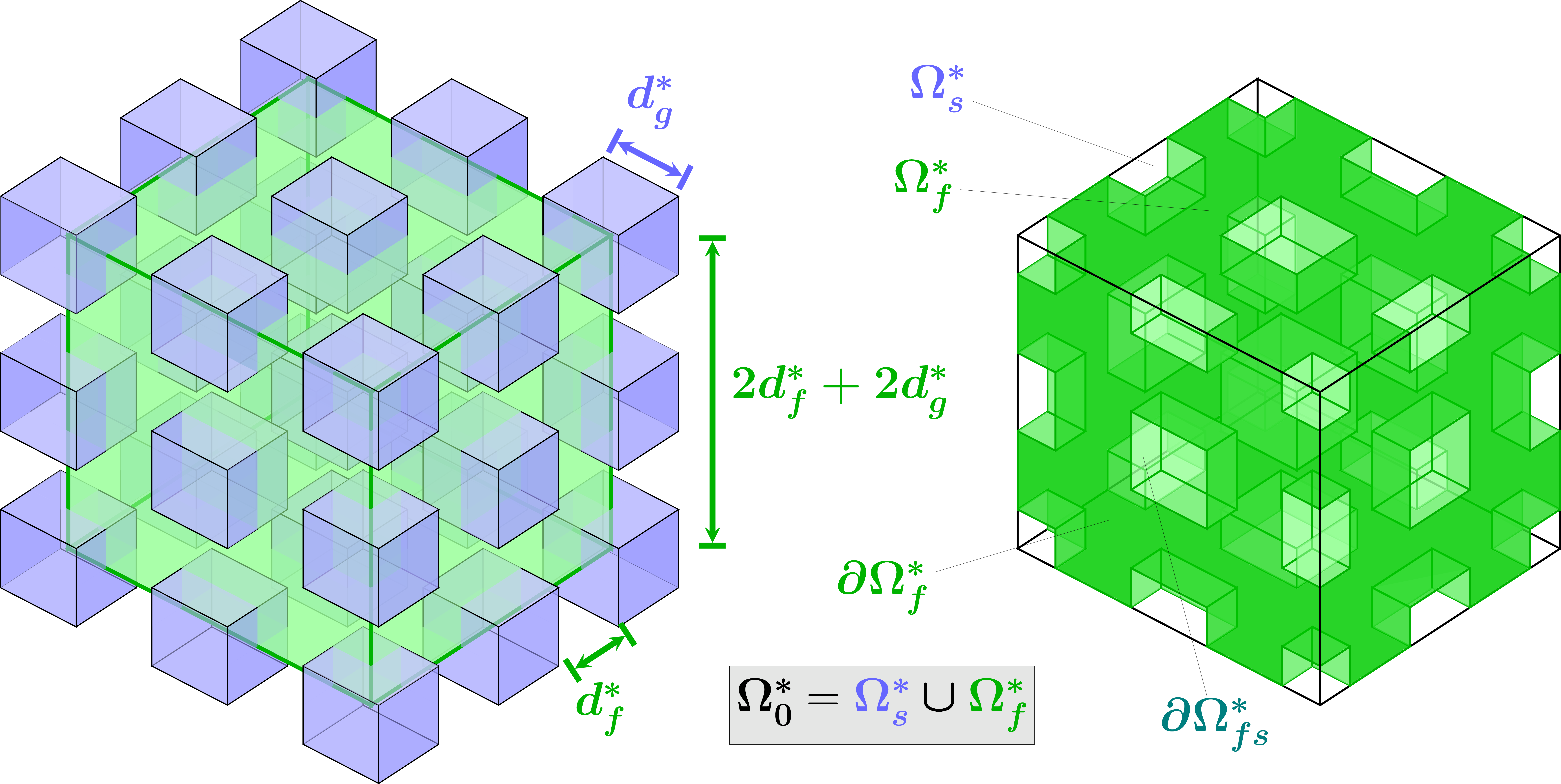}
    \caption[Schematic of the microscopic structure of the porous substrate.]{Schematic of the microscopic structure of the porous substrate. On the left side, a cubic REV (green) of volume $\Omega_0^\ast=8(d_g^\ast+d_f^\ast)^3$ is placed within an array of uniformly-spaced solid grains (blue cubes) of size $d_g^\ast$ and inter-grain distance $d_f^\ast$. On the right, the volume of fluid enclosed by the REV is shown (green). The fluid and solid volume fractions, $\Omega_f^\ast$ and $\Omega_s^\ast$, are also drawn along with the fluid-solid interface surface $(\p\Omega_{fs})^\ast$ and the portion of the external surface of the REV occupied by the fluid $(\p\Omega_f)^\ast$.}
    \label{fig:porous_structure}
\end{figure}

The focus is on a two-dimensional, steady air flow over the top flat surface of an isotropic and uniform porous substrate. The substrate lies above a solid impermeable surface and is saturated with air. The governing equations for the fluid phase are derived by volume averaging the compressible Navier-Stokes equations \cite{Whitaker_1969,Bachmat_Bear_1986,Bear_Bachmat_1990,Whitaker_1998,Sorek_Levi-Hevroni_Levy_Ben-Dor_2005}. The averaged \textit{macroscopic} fluid quantities result from an upscaling process as the \textit{microscopic} quantities are smoothed by applying a spatial filter $\overline{m}^\ast$ over a representative elementary volume (REV) that encloses portions of the fluid domain and the solid matrix. All dimensional quantities are denoted by the superscript $\ast$. When the top-hat filter $\overline{m}^\ast=1/\Omega_0^\ast\, \left[\si{\meter^{-3}}\right]$ \cite{Quintard_Whitaker_1994,Breugem_Boersma_2005} is used (where $\Omega_0^\ast$ is the total volume of the REV), the intrinsic volume average of the fluid phase within the REV is $\fluid{\cdot}=1/\Omega_0^\ast\int_{\Omega_0^\ast}[\cdot]\d\Omega^\ast$. The ratio of the fluid volume to the total volume of the REV defines the volume porosity $\theta_f=\Omega_f^\ast/\Omega_0^\ast$. The surface porosity is the ratio $\theta_{\p ff}=(\p\Omega_f)^\ast/(\p\Omega_0)^\ast$, where $(\p\Omega_f)^\ast$ is the portion of the external surface of the REV wetted by the fluid phase and $(\p\Omega_0)^\ast$ is the total external surface of the REV. A porous substrate with the ordered microstructure is considered (refer to figure \ref{fig:porous_structure}). A cellular filter, which results from the double convolution of the top-hat filter, is employed \cite{Quintard_Whitaker_1994,Breugem_Boersma_2005,Breugem_Boersma_Uittenbogaard_2006}. The averaged macroscopic quantities are defined at the centroid of the REV and vary continuously across an interface of finite thickness at the top boundary of the substrate. To simplify the notation, the normalized filter $m=\overline{m}^\ast\Omega_0^\ast$ and the ratio $\Sigma_{fs}^\ast=(\p\Omega_{fs})^\ast/\Omega_0^\ast\, \left[\si{\per\meter}\right]$ are introduced, where $(\p\Omega_{fs})^\ast$ is the fluid-solid interface surface within the REV \cite{Bachmat_Bear_1986}. The averaging operators read 
\begin{subequations}\label{eq:averaging_filters_def}
\begin{equation}\label{eq:averaging_filters_def_volume}
    \displaystyle\fluid{\left[\cdot\right]} = \frac{1}{\Omega_0^\ast}\int_{\Omega_0^\ast}m\left[\cdot\right]\,\d\Omega^\ast,
\end{equation}
\begin{equation}\label{eq:averaging_filters_def_interphase}
    \qquad \displaystyle\averageinterphasefluxfs{\left[\cdot\right]n_j} \Sigma_{fs}^\ast = \frac{1}{\Omega_0^\ast}\int_{\left(\p\Omega_{fs}\right)^\ast}m\,\left[\cdot\right] n_j\,\d\left(\p\Omega^\ast\right),
\end{equation}
\end{subequations}
and the quantities are decomposed as the sum of their volume-averaged and fluctuating parts, $\left[\cdot\right] = \fluid{\left[\cdot\right]} + \widetilde{\left[\cdot\right]}$, with $\langle\widetilde{\left[\cdot\right]}\rangle_f=0$ \cite{Gray_1975}. The flow is described in a Cartesian frame of reference, where $x^\ast$ and $y^\ast$ define the streamwise and wall-normal coordinates, respectively. The intrinsic average of the gradient (or the divergence) of a tensor $G_{ijk}^\ast$ of dimension three or lower \cite[eq. 2.3.29]{Bear_Bachmat_1990} takes the form
\begin{equation}\label{eq:average_gradient_divergence}
    \theta_f\fluid{\frac{\p G_{ijk}^\ast}{\p x_\sigma^\ast}} = \frac{\p}{\p x_\sigma^\ast}\left(\theta_f\fluid{G_{ijk}^\ast}\right) + \int_{\p\Omega_{fs}}\overline{m}^\ast\, G_{ijk}^\ast n_\sigma\d\left(\p\Omega^\ast\right) ,  
\end{equation}
where $n_i$ are the versors normal to the fluid-solid interface,and $\sigma=i$ for the gradient ($\sigma=j$ for the divergence). The pressure $p^\ast$, the density $\rho^\ast$, the temperature $T^\ast$,and the velocity components $u_i^\ast$ are also introduced. Provided that $\p^2p^\ast/\p x_j^{\ast2}=0$ within the REV, the average of the pressure gradient can be expanded as \cite[eq. 2.3.48]{Bear_Bachmat_1990} 
\begin{equation}\label{eq:average_pressure_gradient}
    \theta_f\fluid{\frac{\p p^\ast}{\p x_i^\ast}} = \theta_f \mathcal{T}_{ij}\frac{\p\fluid{p}^\ast}{\p x_j^\ast} + \int_{\p\Omega_{fs}} \overline{m}^\ast\,\widetilde{x}_i^\ast\frac{\p \widetilde{p}^\ast}{\p x_j^\ast}n_j \d\left(\p\Omega^\ast\right),
\end{equation}
where $\mathcal{T}_{ij}$ is the non-dimensional tortuosity tensor, which is $\left(\theta_{\p ff}/\theta_f\right) \delta_{ij}$ in an isotropic medium \cite{Bachmat_Bear_1986,Bear_Bachmat_1990}. 

\subsection{Governing equations}\label{sec:gov_eqs_porous_base_flow}
We consider the steady, two-dimensional form of the equations for mass conservation and balance for the streamwise and wall-normal momentum and the enthalpy. The solid matrix is assumed to be rigid. The divergence operator \eqref{eq:average_gradient_divergence} is applied to the steady, two-dimensional mass conservation equation. By neglecting the mechanical-dispersion terms $\fluid{\widetilde{\rho}^\ast \widetilde{u}^\ast}$ and assuming zero-mass transport across the impermeable fluid-solid interfaces within the REV, one finds \cite{Bear_Bachmat_1990}
\begin{equation}\label{eq:mass_conservation_dim}
    \frac{\p}{\p x_j^\ast}\left(\theta_f\fluid{\rho^\ast} \fluid{u_j^\ast}\right) = 0,
\end{equation}
where the volume porosity is, in general, a smooth function of $x^\ast$ and $y^\ast$. The momentum balance takes the form
\begin{equation}\label{eq:momentum_balance_dim_porous}
    \theta_f\fluid{\rho^\ast}\fluid{u_j^\ast}\frac{\p\fluid{u_i^\ast}}{\p x_j^\ast} + \theta_f\mathcal{T}_{ij}\frac{\p\fluid{p^\ast}}{\p x_j^\ast} = \theta_f\fluid{\frac{\p\tau_{ij}^\ast}{\p x_j^\ast}} - \averageinterphasefluxfs{\widetilde{x}_i^\ast\frac{\p \widetilde{p}^\ast}{\p x_j^\ast}n_j}\Sigma_{fs}^\ast.
\end{equation}
The solid and fluid phase velocities are zero at the solid-fluid interface and the divergence of the shear stresses $\tau_{ij}^\ast$ expands as \cite[eq. 2.6.32]{Bear_Bachmat_1990}
\begin{multline}\label{eq:momentum_stresses_expanded_porous}
    \theta_f\fluid{\frac{\p\tau_{ij}^\ast}{\p x_j^\ast}} = \frac{\p}{\p x_j^\ast}\left[\fluid{\mu^\ast}\left(\frac{\p \left(\theta_f \fluid{u_i^\ast}\right)}{\p x_j^\ast} + \frac{\p \left(\theta_f \fluid{u_j^\ast}\right)}{\p x_i^\ast}\right)\right] + \\
    + \frac{\p}{\p x_i^\ast}\left[\fluid{\lambda^\ast}\frac{\p\left(\theta_f\fluid{u_k^\ast}\right)}{\p x_k^\ast}\right] + \averageinterphasefluxfs{\mu^\ast\left(\frac{\p u_i^\ast}{\p x_j^\ast}+\frac{\p u_j^\ast}{\p x_i^\ast}\right)}\Sigma_{fs}^\ast + \\
    - \frac{\fluid{\lambda^\ast}}{\theta_f}\frac{\p\left(\theta_f\fluid{u_k^\ast}\right)}{\p x_k^\ast}\frac{\p\theta_f}{\p x_i^\ast},
\end{multline}
where $\lambda^\ast$ is the second coefficient of viscosity and the last term is obtained by employing the identity $\p\theta_f/\p x_i^\ast = \averageinterphasefluxfs{-n_i}\Sigma_{fs}^\ast$. The surface integrals on the right-hand side of equations \eqref{eq:momentum_balance_dim_porous} and \eqref{eq:momentum_stresses_expanded_porous} are often modeled using a permeability tensor \cite{Bachmat_Bear_1986,Whitaker_1986} and a Forchheimer tensor \cite{Whitaker_1996}. The latter is a nonlinear correction to Darcy's law which arises at large microscopic Reynolds numbers \cite{Whitaker_1996,Breugem_Boersma_2005}. While the departure from the linear Darcian regime is very well known and has been widely reported in numerical and laboratory experiments, the mathematical modelling and the physical nature of the Forchheimer correction are a matter of current research. Nonlinear corrections with cubic, quadratic and power-law dependence on the velocity have been reported for different ranges of the microscopic Reynolds numbers on pre-transitional fluid flows inside porous media \cite{Lasseux_Valdes-Parada_2017,Khalifa_Pocher_Tilton_2020}. The surface integral is modeled as the sum of a linear and a quadratic drag \cite{Barrere_Gipouloux_Whitaker_1992,Whitaker_1996,Breugem_Boersma_Uittenbogaard_2006} 
\begin{multline}\label{eq:darcy_forchheimer_expansion}
    \averageinterphasefluxfs{-\widetilde{x}_i^\ast\frac{\p \widetilde{p}^\ast}{\p x_j^\ast}n_j + \mu^\ast\left(\frac{\p u_i^\ast}{\p x_j^\ast}+\frac{\p u_j^\ast}{\p x_i^\ast}\right)n_j}\Sigma_{fs}^\ast =  -\fluid{\mu^\ast}\theta_f^2 \left(K^{\ast-1}\right)^{ij}\fluid{u_j^\ast} + \\ - \fluid{\rho^\ast}\theta_f^2\left[\left(K^{\ast-1}\right)^{ik}\left(c_{F,kjl}^\ast\fluid{u_l^\ast}\right)\right]\fluid{u_j^\ast},
\end{multline}
where the inverse of the permeability tensor $K_{ij}^\ast$ and the Forchheimer tensor $c_{F,kjl}^\ast$ have been introduced. The coefficients of the Darcy and Forchheimer terms are often modeled using the Kozeny-Carman and Ergun relations \cite{Whitaker_1996,Breugem_Boersma_Uittenbogaard_2006}, that is
\begin{equation}\label{eq:permeability_forchheimer_definition}
    K_{ij}^\ast = \frac{\theta_f^3}{\left(1-\theta_f\right)^2} \frac{d_{g}^{\ast2}}{A}\delta_{ij} \mbox{ and } c_{F,ijk}^\ast = \frac{\theta_f}{1-\theta_f}\frac{d_{g}^\ast}{B} \delta_{ijk},
\end{equation}
where the grain size $d_g^\ast=d_g^\ast(x^\ast,y^\ast,d_{g0}^\ast)$ is assumed to be a smooth function of $x^\ast$, $y^\ast$ and the reference value $d_{g0}^\ast$. $A$ and $B$ are empirical coefficients \cite{Whitaker_1996}. Similar forms of the Kozeny-Carman relation of the type $K^\ast = C^\ast\theta_f^m/(1-\theta_f)^n$, where $C^\ast$ is a dimensional parameter and $m$ and $n$ are positive real constants, have also been used \cite{Costa_2006}. The Forchheimer coefficient $c_{F,ijk}^\ast$ is sometimes modeled as an exponential function of $K^\ast$ \cite{Innocentini_Sepulveda_Ortega_2006,Wedin_Cherubini_2016}. Although the closure model \eqref{eq:darcy_forchheimer_expansion} may no longer hold in the compressible regime, where the seepage velocity can be as large as $10^2\,\si{\meter\per\second}$ for high porosity \cite{Mironov_Maslov_Poplavskaya_Kirilovskiy_2015}, equations \eqref{eq:darcy_forchheimer_expansion} and \eqref{eq:permeability_forchheimer_definition} are used with the incompressible values $A=180$ and $B=100$ because of the lack of experimental data. Some authors \cite{Nield_1991,Whitaker_1996,Tilton_Cortelezzi_2015} have argued that the inertial terms should be removed from the equations when a quadratic Forchheimer correction is included, while others \cite{Ochoa-Tapia_Whitaker_1995a,Breugem_Boersma_Uittenbogaard_2006,Tilton_Cortelezzi_2008} have pointed out that they are not negligible at the free fluid-porous interface or in compressible porous media flows \cite{Emanuel_Jones_1968,Shreeve_1968,Nield_1994}. The coexistence of the inertial terms and the Forchheimer terms in the momentum equation is still a matter of debate. We shall retain both terms and adopt the volume-averaging framework of \cite{Breugem_Boersma_2005}. 

Applying the operators \eqref{eq:average_gradient_divergence} to the static enthalpy balance yields
\begin{multline}\label{eq:enthalpy_balance_dim}
    \theta_f\fluid{\rho^\ast}\fluid{u_j^\ast}\frac{\p\fluid{T^\ast}}{\p x_j^\ast} = \frac{\fluid{\mu^\ast}}{\theta_f}\left[\frac{\p\left(\theta_f\fluid{u_i^\ast}\right)}{\p x_j^\ast}\right]^2 + \frac{\fluid{\lambda^\ast}}{\theta_f}\left[\frac{\p\left(\theta_f\fluid{u_k^\ast}\right)}{\p x_k^\ast}\right]^2 +\\ 
    + \frac{\fluid{\mu^\ast}}{\theta_f}\frac{\p\left(\theta_f\fluid{u_i^\ast}\right)}{\p x_j^\ast}\frac{\p\left(\theta_f\fluid{u_j^\ast}\right)}{\p x_i^\ast}
    + \frac{\p}{\p x_j^\ast}\left(\theta_f\fluid{k^\ast}\mathcal{T}_{jk}\frac{\p \fluid{T^\ast}}{\p x_k^\ast}\right),
\end{multline}
where $k^\ast$ is the thermal conductivity. The perfect gas equation becomes $\fluid{p^\ast}= \fluid{\rho^\ast} \mathcal{R}^\ast \fluid{T^\ast}$, where $\mathcal{R}^\ast = 287.05\,\si{\joule\per\kilo\gram\per\kelvin}$ is the specific gas constant of air. We shall assume that the microscopic temperature gradients of the fluid phase at the surface of the solid grains are negligible and that the fluid and solid phases are in local thermal equilibrium (LTE) $\fluid{T^\ast}=\solid{T^\ast}$ \cite[eq. 2.6.121]{Bear_Bachmat_1990}. It is important to note that although the LTE hypothesis has been invoked in the context of steady porous media flows with moderate $\theta_f$ \cite[e.g.][]{Nield_Kutsenov_2003}, it is not valid in general. \cite{Celli_Rees_Barletta_2010} computed the evolution of the temperature profiles of the fluid and solid phases over a flat plate immersed in a uniform porous medium and showed that the LTE assumption is not valid near the leading edge. Thermal equilibrium is only achieved at moderate downstream locations when the conductivity of the fluid phase is much larger than that of the solid phase $k_f^\ast/k_s^\ast\gg 1$ or when the volume porosity is very high $\theta_f\to1$. A similar flow configuration was investigated by \cite{Papalexandris_2023b} for the case of a iron-made, regular-microstructure solid matrix ($k_s^\ast=52\,\si{\watt\per\meter\per\kelvin}$) saturated with water and air. The temperature profiles for the fluid and solid phases differed significantly at short downstream distances $x^\ast/d_{g0}^\ast=O(1)$. One must exercise caution when invoking the LTE hypothesis as local non-equilibrium may occur, especially in the vicinity of the leading edge.

We shall assume the LTE hypothesis to be valid in the porous layer because we expect the velocity to be very small (and the static temperature to be comparable to the adiabatic recovery temperature) below the porous-fluid interface. The static enthalpy balance of the fluid phase and the internal energy balance of the solid phase then reduce to the thermal conduction and thermal convection terms \cite{Celli_Rees_Barletta_2010}. When an adiabatic condition is imposed underneath, the temperature across the substrate is almost uniform and the interphasial heat exchange is negligible.

The fluid flow is uniform away from the substrate because the streamwise pressure gradient is null. A thin boundary layer forms above the fluid-porous interface. The velocity components, the density, the temperature and the transport coefficients are normalized by their free-stream values, denoted by the subscript $\infty$. The macroscopic pressure $\fluid{p^\ast}$ is scaled by $\rho_\infty^\ast U_\infty^{\ast2}$, where $U_\infty^\ast$ is the free-stream velocity. The free-stream Reynolds number is $\Reynolds = \rho_\infty^\ast U_\infty^\ast L^\ast/\mu_\infty^\ast\gg1$, where $\mu_\infty^\ast$ is the free-stream dynamic viscosity and $L^\ast$ is a characteristic large streamwise length. The Darcy number, based on the characteristic grain size $d_{g0}^\ast$, is $\Darcy = d_{g0}^{\ast2}/L^{\ast2}\ll 1$ and the free-stream Mach number is $\Mach = U_\infty^\ast/c_\infty^\ast = \order{1}$,
where $c_\infty^\ast=(\gamma\mathcal{R}^\ast T_\infty^\ast)^{1/2}$ is the speed of sound in the free stream and $\gamma=1.4$ the heat capacity ratio. The dynamic viscosity is modeled using Sutherland's law
\begin{equation}
    \frac{\mu^\ast}{\mu_{\infty}^\ast} = \left(\frac{T^\ast}{T_{\infty}^\ast}\right)^{3/2}\frac{\chi^\ast+T_{\infty}^\ast}{\chi^\ast+T^\ast}, 
\end{equation}
where $\chi^\ast=110\,\si{\kelvin}$ is the Sutherland temperature. The parabolic character of the solution obtained by \cite{Vafai_Kim_1990} resulted from the concurrent contribution of the Brinkman, Forchheimer and advective terms, all of which need to be retained while performing a rigorous asymptotic analysis. The Forchheimer term cannot satisfy the boundary-layer approximation when scaled by a factor $\rho_\infty^\ast U_\infty^\ast/d_{g0}^\ast\gg 1$ \cite{Tsiberkin_2018b}. The Darcy and Forchheimer terms are a parameterization of the surface integrals in \eqref{eq:darcy_forchheimer_expansion} and appropriate scaling must be applied to the integrating functions themselves. The characteristic length of the REV $\Sigma_{fs}^{\ast-1}$ scales as $d_{g}^\ast$, the velocity scales as $U_\infty^\ast$ \cite{Barrere_Gipouloux_Whitaker_1992,Harter_Martinez_Poser_Weigand_Lamanna_2023} and the microscopic pressure fluctuations $\widetilde{p}^\ast$ scale as $\mu_\infty^\ast U_\infty^\ast/d_{g}^\ast$ \cite{Whitaker_1996}. 

\subsection{Porous-free fluid interface}
\begin{figure}[ht]
    \centering
    \includegraphics[width=\linewidth]{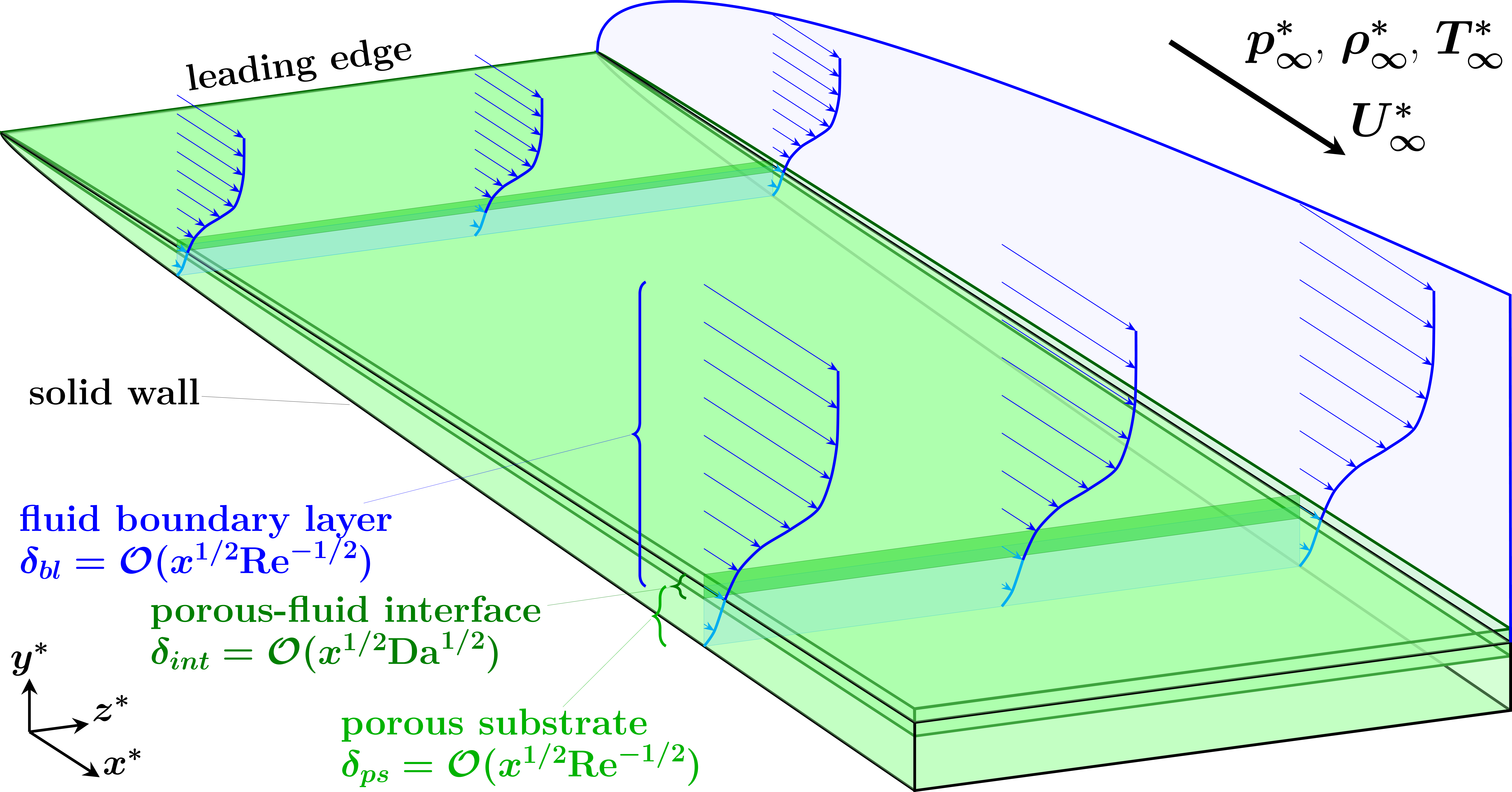}
    \caption[Schematic of the self-similar flow over an isotropic porous substrate.]{Schematic of the self-similar flow. A boundary layer of thickness $\delta_{bl}$ (blue) evolves over an isotropic porous substrate of streamwise-increasing thickness $\delta_{ps}$ (green). The figure is not to scale.}
    \label{fig:self-similar_domain}
\end{figure}
A schematic of the self-similar, coupled porous substrate-boundary layer flow is shown in figure \ref{fig:self-similar_domain}. As in classic boundary-layer theory \cite{Stewartson_1964}, the flow properties are expected to vary in a region of thickness $\order{\Reynolds^{-1/2}}$ near the top surface of the substrate. From this point onwards, the averaging operators $\fluid{\cdot}$ are omitted for clarity, with no risk of ambiguity regarding volume averaging. The limiting form of equations \eqref{eq:mass_conservation_dim}, \eqref{eq:momentum_balance_dim_porous} and \eqref{eq:enthalpy_balance_dim} for $\Reynolds\gg1$ and $\Darcy\ll1$ is parabolic in $x$ and reads \cite{van-Dyke_1975}
\begin{subequations}\label{eq:coupled_por_BL}
\begin{align}
    \frac{\p}{\p x}\left(\frac{\theta_f u}{T}\right) + \frac{\p}{\p y}\left(\frac{\theta_f v}{T}\right) &=0, \label{eq:coupled_por_BL_continuity} \\
    \frac{\theta_f u}{T}\frac{\p u}{\p x} + \frac{\theta_f v}{T}\frac{\p u}{\p y} &= \frac{\p}{\p y}\left[\mu\frac{\p \left(\theta_f u\right)}{\p y}\right] - \frac{\theta_f^2}{\widehat{\kappa}_p^2}\left(\frac{\mu u}{K} + \frac{\theta_f \widehat{c}_F}{K^{1/2}}\frac{u^2}{T}\right), \label{eq:coupled_por_BL_sw_momentum}\\
    \frac{\theta_f u}{T} \frac{\p T}{\p x} + \frac{\theta_f v}{T} \frac{\p T}{\p y} &= \frac{\mu}{\theta_f}\left[\frac{\p\left(\theta_f u\right)}{\p y}\right]^2 + \frac{\p}{\p y}\left(\theta_{\p ff} \frac{\mu}{\Prandtl}\frac{\p T}{\p y}\right), \label{eq:coupled_por_BL_enthalpy}
\end{align}
\end{subequations}
where the streamwise pressure gradient is absent and the perfect gas equation is $\rho T=1$. A self-similar solution can not be retrieved in the general case where the Forchheimer coefficient is a function of a streamwise-varying grain diameter $d_g^\ast=d_{g0}^\ast\left(2x\right)^{1/2}$. It can only be obtained if the permeability $K^\ast$ is a function of the streamwise-varying grain diameter $d_g^\ast=d_{g0}^\ast\left(2x\right)^{1/2}$ and the Forchheimer coefficient $\widehat c_F$ is a function of a constant grain diameter $d_g^\ast=d_{g0}^\ast$. $d_{g0}^\ast$, i.e. $\widehat{c}_F=\widehat{c}_F(\theta_f,d_{g0}^\ast)$. When the low-speed, incompressible closure model \eqref{eq:darcy_forchheimer_expansion} is considered, this choice is equivalent to a local-similarity assumption. Here $K=\theta_f^3/[(1-\theta_f)^2A] =\order{1}$, $\widehat{c}_F= A^{1/2}\Reynolds\Darcy /(B\Darcy^{1/2}\theta_f^{3/2}) = \order{1}$, $\widehat{\kappa}_p^2=\Reynolds\Darcy (d_g^\ast/d_{g0}^\ast)^2=\order{1}$ \cite{Tsiberkin_2016}, $A=\order{1}$ and $B=O(\Darcy^{-1/2})$. The Prandtl number of air is $\Prandtl=c_p^\ast\mu_\infty^\ast/k_\infty^\ast=0.71$ and $k(T)=\mu(T)$ \cite{Stewartson_1964}. A detailed derivation of the self-similar momentum equation and the Darcy-Forchheimer term is found in appendix \ref{app:forchheimer}. 

A no-slip, no-penetration condition is imposed at the bottom solid wall of the substrate. The boundary conditions read
\begin{equation}\label{eq:temperature_boundary_conditions}
\begin{array}{cc}
    \displaystyle u\left(x,y_w\right) = 0, & \displaystyle u\left(x,\infty\right) = 1, \\
    \displaystyle \frac{\p T}{\p y}\left(x,y_w\right) = 0, & \displaystyle T\left(x,\infty\right) = 1 ,
\end{array}
\end{equation}
where the subscript $w$ denotes the bottom solid wall. The top surface of the porous substrate is flat and located at a height $y_{int}^\ast$. 

\begin{figure}[ht]
    \centering
    {\includegraphics[width=0.49\linewidth]{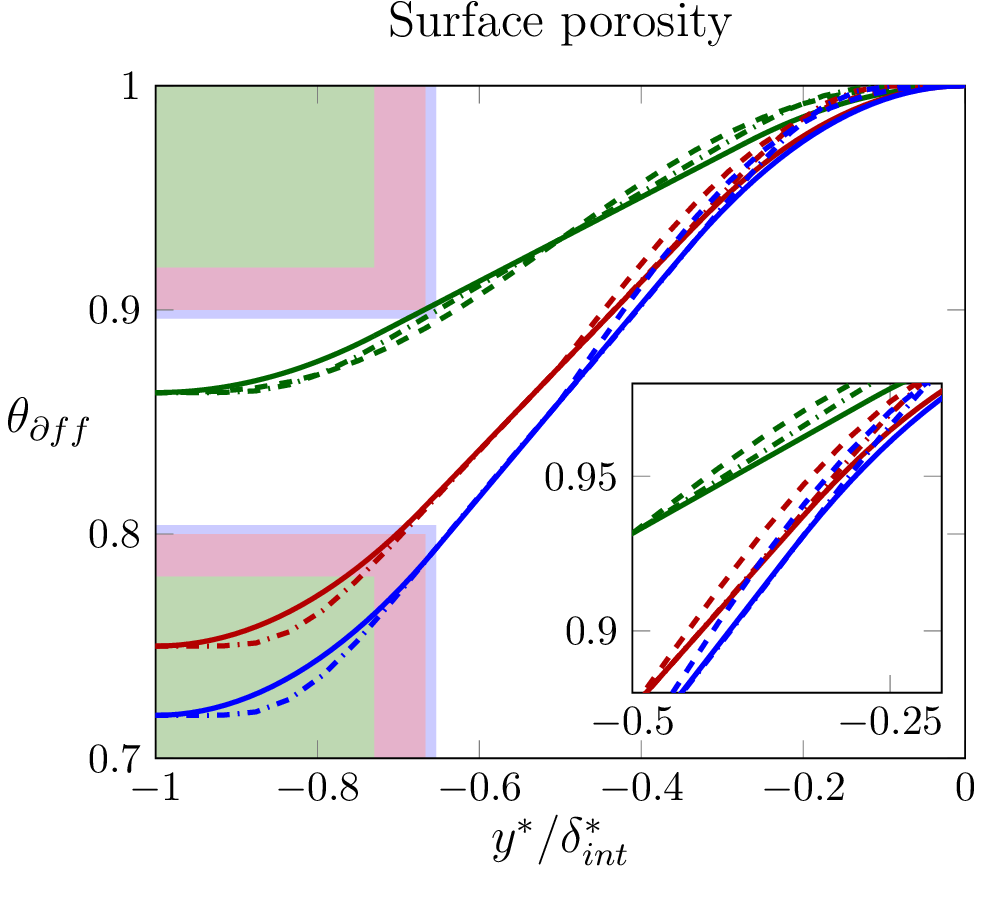}
    \includegraphics[width=0.49\linewidth]{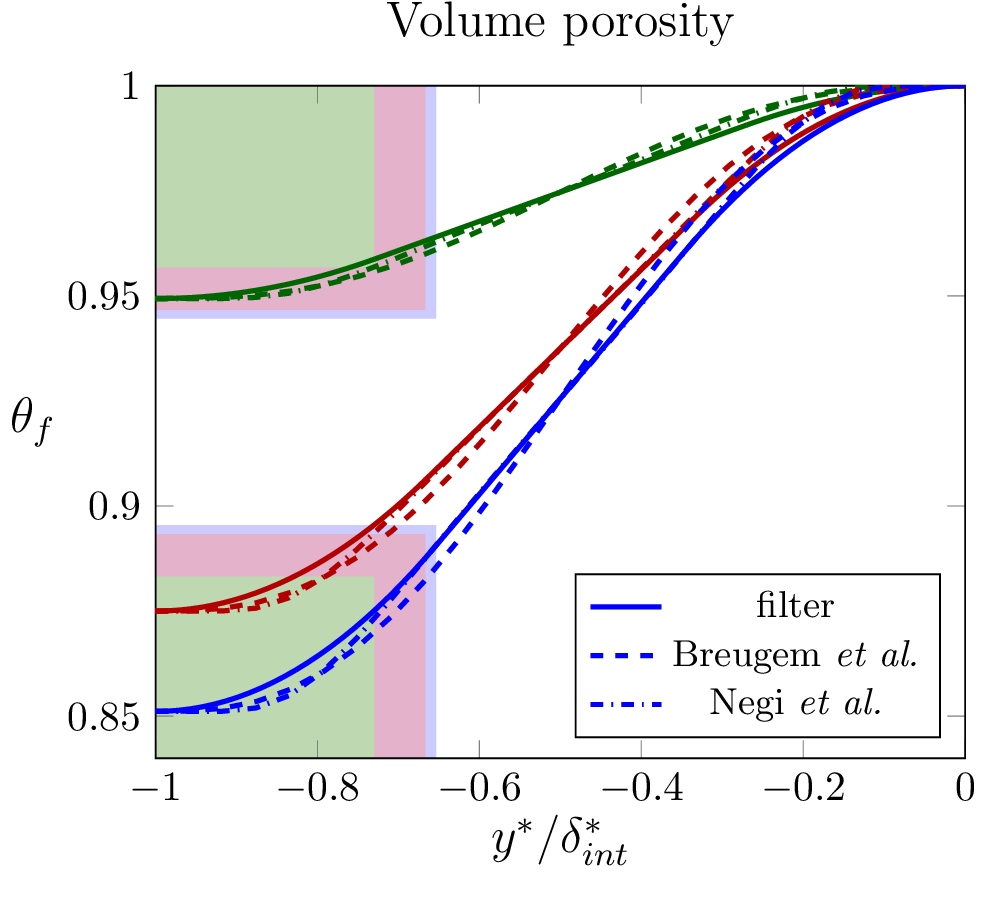}}
    \caption[Distribution of the volume and surface porosity across the interface.]{Distribution of the volume and surface porosity across the interface. Comparison of the model equations of \cite{Breugem_Boersma_2005} (dashed curves) and \cite{Negi_Mishra_Skote_2015} (dash dot curves) \eqref{eq:negi_model} with the results obtained by computing the convolution integral of the cellular filter across the porous-free fluid interfacial layer of thickness $\delta_{int}^\ast$ \eqref{eq:interfacial_porosity_general} (solid curves) for $Q=0.37$ (green), $Q=0.5$ (red) and $Q=0.53$ (blue). Green, red and blue solid grains of size $d_{g}^\ast = \delta_{int}^\ast Q/(1+Q)$ are drawn at the bottom side of the interface for comparison.}
    \label{fig:interfacial_model}
\end{figure}
The thickness of the interface is $\delta_{int}=O(\Darcy^{1/2})$ \cite{Goharzadeh_Khalili_Jorgensen_2005} and is therefore comparable to that of the boundary layer because $\delta_{int}/\delta_{bl}=O(\Reynolds^{1/2}\Darcy^{1/2})=\order{1}$. The volume and surface porosity vary smoothly across the interface. Their uniform values below the interface are $\theta_{fp}=1-Q^3$ and $(\theta_{\p ff})_p = 1-Q^2$, respectively, where $Q=d_g^\ast/(d_f^\ast+d_g^\ast)$. Here, $d_f^\ast=d_f^\ast(x^\ast,y^\ast,d_{f0}^\ast)$, where $d_{f0}^\ast$ is a characteristic constant. The interfacial thickness is $\delta_{int}^\ast=d_g^\ast\,(1+Q)/Q$. The distribution of $\theta_f$ and $\theta_{\p ff}$ is obtained analytically by sweeping the averaging operator \eqref{eq:averaging_filters_def} across the interfacial region \cite{Breugem_2005}. Repeating this procedure for arbitrary $d_g^\ast$ and $d_f^\ast$ yields the analytical piecewise-polynomial curve shown in figure \ref{fig:interfacial_model} (solid curves)
\begin{subequations}\label{eq:interfacial_porosity_general}
\begin{equation}
    \theta_{\p ff}\left(\frac{y^\ast}{\delta_{int}^\ast}\right) = 1 \mbox{ for }\frac{y^\ast}{\delta_{int}^\ast}\geq 0, 
\end{equation}
\begin{multline}
    \theta_{\p ff}\left(\frac{y^\ast}{\delta_{int}^\ast}\right) = 1 - \frac{Q}{\left(d_f^\ast+d_g^\ast\right)^2}\int_{-d_g^\ast-d_f^\ast}^{-d_g^\ast-d_f^\ast -y}\left(d_g^\ast+d_f^\ast+\breve{y}^\ast\right)\d \breve{y}^\ast = \\ 
    = 1 - \frac{Q}{2}\left(1+Q\right)^2\left(\frac{y^{\ast}}{\delta_{int}^{\ast}}\right)^2 \mbox{ for } -\frac{Q}{1+Q} \leq \frac{y^\ast}{\delta_{int}^\ast} < 0, 
\end{multline}
\begin{multline}
    \theta_{\p ff}\left(\frac{y^\ast}{\delta_{int}^\ast}\right) = 1 - \frac{Q}{\left(d_f^\ast+d_g^\ast\right)^2}\int_{-d_f^\ast-2d_g^\ast-y^\ast}^{-d_g^\ast-d_f^\ast-y^\ast}\left(d_g^\ast+d_f^\ast+\breve{y}^\ast\right)\d \breve{y}^\ast = \\
    = 1 + Q^3\left[\frac{1}{2} + \frac{1+Q}{Q}\frac{y^\ast}{\delta_{int}^\ast}\right] \mbox{ for } -\frac{1}{1+Q}\leq \frac{y^\ast}{\delta_{int}^\ast}<-\frac{Q}{1+Q}, 
\end{multline}
\begin{multline}
    \theta_{\p ff}\left(\frac{y^\ast}{\delta_{int}^\ast}\right) = 1 - \frac{Q}{\left(d_f^\ast+d_g^\ast\right)^2}\left[\int_{-d_g^\ast-d_f^\ast}^{-2d_g^\ast-2d_f^\ast-y}\left(d_g^\ast + d_f^\ast+\breve{y}^\ast\right)\d \breve{y}^\ast \right. + \\
    + \int_{0}^{-d_f^\ast-2d_g^\ast-y^\ast}\left(d_g^\ast+d_f^\ast+\breve{y}^\ast\right)\d \breve{y}^\ast + \left.\int_{-d_g^\ast-d_f^\ast-y^\ast}^{0}\left(d_f^\ast+d_g^\ast-\breve{y}^\ast\right)\d \breve{y}^\ast\right] = \\
    = 1 + Q\left[\frac{\left(1+Q^2\right)}{2} + \left(1+Q\right)^2\frac{y^\ast}{\delta_{int}^\ast} + \frac{\left(1+Q\right)^2}{2}\left(\frac{y^\ast}{\delta_{int}^\ast}\right)^2\right] \\
    \mbox{ for } -1 \leq \frac{y^\ast}{\delta_{int}^\ast}<-\frac{1}{1+Q},
\end{multline} 
\begin{equation}
    \theta_{\p ff}\left(\frac{y^\ast}{\delta_{int}^\ast}\right) = 1-Q^2 \mbox{ for } \frac{y^\ast}{\delta_{int}^\ast}<-1.
\end{equation}
\end{subequations}
\cite{Breugem_Boersma_2005} approximated \eqref{eq:interfacial_porosity_general} with a fifth-order polynomial. Instead, an interpolating exponential function is used in this work \cite{Negi_Mishra_Skote_2015}
\begin{equation}\label{eq:negi_model}
    \frac{\theta_f(\widetilde{y})-\theta_{fp}}{1-\theta_{fp}} = \frac{\theta_{\p ff}(\widetilde{y})-(\theta_{\p ff})_p}{1-(\theta_{\p ff})_p} = \left[\displaystyle 1+\exp\left(\frac{C}{\widetilde{y}}+\frac{C}{\widetilde{y}+1}\right)\right]^{-1}
\end{equation}
where $\widetilde{y}=\left(y^\ast-y_{int}^\ast\right)/\delta_{int}^\ast$, $C=0.75$. As shown in figure \ref{fig:interfacial_model}, the agreement between the analytical piecewise curve and the exponential model \eqref{eq:negi_model} is excellent. 

\subsection{Self-similar solution}
A similarity solution of the differential system \eqref{eq:coupled_por_BL} is sought in this section. First, the Dorodnitsyn-Howarth variable \cite{Stewartson_1964}
\begin{equation}
    \ybar = \int_0^y\rho(x,\Breve{y})\d \Breve{y},
\end{equation}
is introduced. A streamfunction $\psi\left(x,\ybar\right)$ is defined so that the continuity equation \eqref{eq:coupled_por_BL_continuity} is satisfied 
\begin{equation}
\begin{array}{cc}
    \displaystyle u=\frac{1}{\theta_f} \frac{\p\psi}{\p\ybar}, & \displaystyle v = -\frac{T}{\theta_f} \frac{\p\psi}{\p x},
\end{array}
\end{equation}
Following a standard procedure in the derivation of the self-similar boundary layer equations, a decomposition of $\psi$ and $\ybar$ is sought in the form
\begin{equation}
\begin{array}{cc}
    \psi\left(x,\ybar\right) = \left(\alpha x\right)^a F\left(x,\eta\right), & \ybar = \left(\alpha x\right)^b \eta,
\end{array}
\end{equation}
where $\alpha$, $a$ and $b$ are real constants. Inspection of \eqref{eq:temperature_boundary_conditions} shows that $T$ cannot be a function of $x$ in homoenthalpic flows and that no similarity solution is possible if an inhomogeneous Neumann condition for the temperature is imposed at the solid wall underneath the substrate. Since the top of the substrate is flat, the wall-normal porosity distribution \eqref{eq:negi_model} is a function of $\eta$ alone when both $\delta_{int}$ and $\delta_{ps}$ are $\order{x^b}$. Following \cite{Tsiberkin_2016,Tsiberkin_2018a,Tsiberkin_2018b}, the grain size and the inter-grain distance are assumed to be smooth functions of $x$, $d_g^\ast=d_{g0}^\ast\left(\alpha x\right)^{c/2}$ and $\widehat{\kappa}_p=\kappa_p\left(\alpha x\right)^c$, where
\begin{equation}\label{eq:control_parameter}
    \kappa_p^2=\Reynolds\Darcy = \frac{\rho_\infty^\ast U_\infty^\ast d_{g0}^{\ast2}}{\mu_\infty^\ast L^\ast} = \order{1}
\end{equation}
is the control parameter defined by \cite{Tsiberkin_2018a}. The parameter distills the effect of the linear Darcy drag for given free-stream conditions and grain size. A similarity solution exists for $a=b=1/2$, $c=1$ and $\alpha=2$. Equations \eqref{eq:coupled_por_BL_sw_momentum} and \eqref{eq:coupled_por_BL_enthalpy} then reduce to a system of coupled ordinary differential equations
\begin{subequations}\label{eq:self-similar_porous}
\begin{align}
    \left(\frac{\mu}{T}F^{\prime\prime}\right)^\prime + F \left(\frac{F^\prime}{\theta_f}\right)^\prime - C_D\mu T\frac{\left(1-\theta_f\right)^2}{\theta_f^2}F^\prime - C_F\frac{1-\theta_f}{\theta_f^2}\left(F^\prime\right)^2 & = 0, \label{eq:self-similar_porous_momentum}\\
    \frac{1}{\Prandtl}\left(\theta_{\p ff}\frac{\mu T^\prime}{T}\right)^\prime + FT^\prime + \frac{\left(\gamma-1\right)}{\theta_f}\frac{\mu}{T}\Mach^2\left(F^{\prime\prime}\right)^2 &= 0, \label{eq:self-similar_porous_enthalpy}
\end{align}
\end{subequations}
where the primes denote differentiation with respect to the similarity variable
\begin{equation}\label{eq:similarity_variable}
    \eta = \frac{\ybar}{\left(2x\right)^{1/2}},
\end{equation}
and the coefficient of the Darcy and the Forchheimer terms are $C_D=A/\kappa_p^2$ and $C_F= A/(B\Darcy^{1/2}) = \order{1}$, respectively. Equations \eqref{eq:self-similar_porous} are subject to the boundary conditions $F=F^\prime=T^\prime=0$ at the bottom solid wall ($\eta$=0) and $F^\prime,T \rightarrow 1$ as the free stream is approached ($\eta \rightarrow \infty$). For the first time, a self-similar form of the governing equations has been derived for compressible flow over a porous substrate with streamwise-increasing permeability, using a closure model with a constant Forchheimer coefficient. Equations \eqref{eq:self-similar_porous} reduce to the compressible Blasius solution \cite{Stewartson_1964} above the interface, where $\theta_f,\theta_{\p ff}= 1$. The velocity components are 
\begin{subequations}\label{eq:porous_base_flow_velocity_components}
\begin{align}
    u\left(\eta\right) &= \frac{F^\prime}{\theta_f}, \label{eq:streamwise_velocity} \\
    v\left(\eta\right) &= \frac{1}{\theta_f\left(2x\right)^{1/2}}\left(-TF+ \eta_cTF^\prime\right), \label{eq:wall-normal_velocity}
\end{align}
\end{subequations}
and $\eta_c = T^{-1}\int_0^\eta T(\Breve{\eta})\d\Breve{\eta}$. The non-dimensional thickness of the fluid boundary layer, the fluid-porous interface and the porous substrate are $\delta_{bl} = O(x^{1/2}\Reynolds^{-1/2})$, $\delta_{int}=O(x^{1/2}\Darcy^{1/2})$ and $\delta_{ps}=O(x^{1/2}\Reynolds^{-1/2})$, respectively. The thickness of the porous substrate can be written explicitly as a function of the distance from the leading edge, i.e. $\delta_{ps}=C_{ps}x^{1/2}\text{\textit{Re}}^{-1/2}$ or $\delta_{ps}=C_{ps}\kappa_p^{-1/2}x^{1/2}\text{\textit{Da}}^{1/2}$, where $C_{ps}=O(1)$ is a design parameter. The interfacial continuity of the surface-averaged tangential velocity $\theta_f u$ derived by \cite{Ochoa-Tapia_Whitaker_1995a} is recovered from \eqref{eq:streamwise_velocity} for small $\kappa_p$ ($\delta_{int}/\delta_{bl}\ll 1$).

The system \eqref{eq:self-similar_porous} is solved by means of a second-order accurate block-elimination algorithm where nonlinearity is treated with the Newton-Raphson method \cite{Cebeci_2002}. A complete description of the numerical procedures is presented in appendix \ref{app:numerical_method}. A uniform grid was employed, and the computations were performed with $N=4000$ and $N=20000$ points to ensure grid independency. The residuals were strictly kept below $10^{-12}$. The interfacial thickness in the $\eta$-space,
\begin{equation}\label{eq:delta_eta_int}
    (\Delta\eta)_{int} = \frac{(\Delta y)_{int}}{T_{av}} = \frac{\kappa_p}{T_{av}}\frac{1+Q}{Q},
\end{equation}
depends on the average temperature of the interfacial region $T_{av}$ and is determined iteratively by solving the governing equations \eqref{eq:self-similar_porous} for tentative $\Delta y_{int}$ and $\Delta \eta_{int}$ until the computed values of $T_{av}$ satisfy \eqref{eq:delta_eta_int}. The values of $(\Delta y)_{int}$, $T_{av}$ and $(\Delta_\eta)_{int}$ are tabulated in table \ref{tab:numerical_parameters_base_flow_porous}. 

%%%%%%%%%%%%%%%%%%%%%%%%%%%%%%%%%%%%%%%%%%%%%%%%%%%%
%%%%%%%%%%%%%%%%%%%%%%%%%%%%%%%%%%%%%%%%%%%%%%%%%%%%
%%%%%%%%%%%%%%%%%%%%%%%%%%%%%%%%%%%%%%%%%%%%%%%%%%%%
%%%%%%%%%%%%%%%%%%%%%%%%%%%%%%%%%%%%%%%%%%%%%%%%%%%%
%%%%%%%%%%%%%%%%%%%%%%%%%%%%%%%%%%%%%%%%%%%%%%%%%%%%

\section{Results}\label{sec4:results}
The results were obtained using physical parameters representative of real supersonic and hypersonic wind-tunnel conditions, as listed in tables \ref{tab:experimental_blayers_porous} and \ref{tab:parameters}. The corresponding numerical parameters, such as the factor $Q$, the interfacial thicknesses $(\Delta y)_{int}$ and $(\Delta\eta)_{int}$ and the surface temperature $T_{av}$, are provided in table \ref{tab:numerical_parameters_base_flow_porous} for reference. 

\begin{table}[!htbp]
\caption[Wind tunnel measurements of compressible boundary layers over impermeable flat plates in the high supersonic regime.]{Wind tunnel measurements of compressible boundary layers over impermeable flat plates. Data are retrieved from \cite[GB02]{Graziosi_Brown_2002}, \cite[{}M01]{Maslov_Shiplyuk_Sidorenko_Arnal_2001}, \cite[R23]{Running_Bemis_Hill_Borg_Redmond_Jantze_Scalo_2023}.}\label{tab:experimental_blayers_porous}
\begin{tabular*}{\textwidth}{@{\extracolsep\fill}lccccccc}
\toprule
Ref. & $\Mach$ & $p_o^\ast\,[\si{\kilo\pascal}]$ & $T_o^\ast\,[\si{\kelvin}]$ & $p_\infty^\ast\,[\si{\kilo\pascal}]$ & $T_\infty^\ast\,[\si{\kelvin}]$ & $T_w^\ast/T_{ad,w}^\ast$ & $\delta_{99}^\ast\,[\si{\milli\meter}]$\textsuperscript{\dag} \\
\midrule
GB02 & $2.98$ & $31$ & $290$ & $0.87$ & $104$ & $1.1$ & $\left[1.8,3.3\right]$ \\
M01 & $5.92$ & $1080$ & $390$ & $0.74$ & $49$ & $1$ & $\left[1.8,2.2\right]$ \\
R23 & $6.1$ & $\left[490,3044\right]$ & $473$ & $56.03$ & $\left[54.25,57.81\right]$ & n.a. & n.a. \\
\bottomrule
\end{tabular*}
\begin{tablenotes}
\small
\item[]\textsuperscript{\dag} Measured between $x=89\,\si{\milli\meter}$ and $x=305\,\si{\milli\meter}$ (GB02) and at $x=96\,\si{\milli\meter}$ (M01).
\end{tablenotes}
\end{table}

%%%%%%%%%%%%%%%%%%%%%%%%%%%%%%%%%%%%%%%%%%%%%%%%%%%%%%%%%%%%%%%%%%%%%%%%%%%%%%%
%%%%%%%%%%%%%%%%%%%%%%%%%%%%%%%%%%%%%%%%%%%%%%%%%%%%%%%%%%%%%%%%%%%%%%%%%%%%%%%
%%%%%%%%%%%%%%%%%%%%%%%%%%%%%%%%%%%%%%%%%%%%%%%%%%%%%%%%%%%%%%%%%%%%%%%%%%%%%%%
\begin{sidewaystable}[!htbp]
\caption[Physical parameters used in chapter 4.]{Physical parameters. The volume and surface porosity in the region below the interface are denoted with $\theta_{fp}$ and $\left(\theta_{\p ff}\right)_p$, respectively.}\label{tab:parameters}
\begin{tabular*}{\textwidth}{@{\extracolsep\fill}cccccccccc}
\toprule%
$\Mach$ & $p_\infty^\ast$ & $p_\circ^\ast$ & $T_\infty^\ast$ & $T_\circ^\ast$ & $U_\infty^\ast$ & $\delta_{99}^\ast$ & $d_{g0}^\ast$ & $\kappa_p^2$ & case \\
- & $\si{\kilo\pascal}$ & $\si{\kilo\pascal}$ & $\si{\kelvin}$ & $\si{\kelvin}$ & $\si{\meter\per\second}$ & $\si{\milli\meter}$ & $\si{\upmu\meter}$ & - & - \\
\midrule
\multirow{2}{*}{0.01} & \multirow{2}{*}{100} & \multirow{2}{*}{100} & \multirow{2}{*}{293} & \multirow{2}{*}{293} & \multirow{2}{*}{3.43} & \multirow{2}{*}{3.3} & 100 & 0.02 & A1 \\
& & & & & & & 200 & 0.09 & A2\\
\multirow{ 2}{*}{3} & \multirow{ 2}{*}{0.82} & \multirow{ 2}{*}{30.12} & \multirow{2}{*}{104} & \multirow{2}{*}{291} & \multirow{2}{*}{611} & \multirow{2}{*}{1.8} & 100 & 0.24 & B1\\
& & & & & & & 200 & 0.96 & B2\\
\multirow{ 2}{*}{6} & \multirow{ 2}{*}{0.76} & \multirow{2}{*}{1200} & \multirow{ 2}{*}{60} & \multirow{ 2}{*}{492} & \multirow{ 2}{*}{927} & \multirow{2}{*}{1.6} & 100 & 1.10 & C1 \\
& & & & & & & 200 & 4.35 & C2 \\
\bottomrule
\end{tabular*}
\begin{tabular*}{\textwidth}{@{\extracolsep\fill}ccccc}
\toprule
$Q$ & $1-Q$ & $ (\theta_{\p ff})_p = 1-Q^2$ & $\theta_{fp} = 1-Q^3$ & $\delta_{int}^\ast/d_{g}^\ast=(1+Q)/Q$ \\
\midrule
0.22 & 0.78 & 0.95 & 0.99 & 5.55 \\
0.37 & 0.63 & 0.86 & 0.95 & 3.70 \\
0.5 & 0.5 & 0.75 & 0.875 & 3 \\
0.53 & 0.47 & 0.72 & 0.85 & 2.89 \\
\bottomrule
\end{tabular*}
\end{sidewaystable}
%%%%%%%%%%%%%%%%%%%%%%%%%%%%%%%%%%%%%%%%%%%%%%%%%%%%%%%%%%%%%%%%%%%%%%%%%%%%%%%
%%%%%%%%%%%%%%%%%%%%%%%%%%%%%%%%%%%%%%%%%%%%%%%%%%%%%%%%%%%%%%%%%%%%%%%%%%%%%%%
%%%%%%%%%%%%%%%%%%%%%%%%%%%%%%%%%%%%%%%%%%%%%%%%%%%%%%%%%%%%%%%%%%%%%%%%%%%%%%%
\begin{sidewaystable}[!htbp]
\caption[Numerical parameters used in chapter 4.]{Numerical parameters used in the computations.}l\label{tab:numerical_parameters_base_flow_porous}
\begin{tabular*}{\textwidth}{@{\extracolsep\fill}ccccccccccc}
\toprule%
$\mathrm{Ma}$ & $d_{g0}^\ast\,[\si{\upmu\meter}]$ & $\kappa_p^2$ & $C_D$ & $C_F$ & $\theta_{fp}$ & $Q$ & $\left(\Delta y\right)_{int}$ & $T_{av}$ & $\left(\Delta\eta\right)_{int}$ & case \\
\midrule
\multirow{4}{*}{0.01} & \multirow{2}{*}{100} & \multirow{2}{*}{0.02} & \multirow{2}{*}{9000} & \multirow{2}{*}{1800} & 0.85 & 0.53 & 0.43 & 1 & 0.43 & \multirow{2}{*}{A1}\\
 & & & & & 0.95 & 0.37 & 0.57 & 1 & 0.57 & \\
 & \multirow{2}{*}{200} & \multirow{2}{*}{0.09} & \multirow{2}{*}{2000} & \multirow{2}{*}{900} & 0.85 & 0.53 & 0.86 & 1 & 0.86 & \multirow{2}{*}{A2}\\
 & & & & & 0.95 & 0.37 & 1.11 & 1 & 1.11 & \\
\midrule
\multirow{4}{*}{3} & \multirow{2}{*}{100} & \multirow{2}{*}{0.24} & \multirow{2}{*}{750} & \multirow{2}{*}{1800} & 0.85 & 0.53 & 1.42 & 2.43 & 0.58 & \multirow{2}{*}{B1}\\
 & & & & & 0.95 & 0.37 & 1.83 & 2.31 & 0.79 & \\
 & \multirow{2}{*}{200} & \multirow{2}{*}{0.96} & \multirow{2}{*}{187.5} & \multirow{2}{*}{900} & 0.85 & 0.53 & 2.83 & 2.32 & 1.22 & \multirow{2}{*}{B2}\\
 & & & & & 0.95 & 0.37 & 3.63 & 2.23 & 1.63 & \\
 \midrule
\multirow{4}{*}{6} & \multirow{2}{*}{100} & \multirow{2}{*}{1.10} & \multirow{2}{*}{163.6} & \multirow{2}{*}{1800} & 0.85 & 0.53 & 3.01 & 6.74 & 0.45 & \multirow{2}{*}{C1}\\
 & & & & & 0.95 & 0.37 & 3.88 & 6.28 & 0.62 & \\
 & \multirow{2}{*}{200} & \multirow{2}{*}{4.35} & \multirow{2}{*}{41.3} & \multirow{2}{*}{900}  & 0.85 & 0.53 & 6.02 & 6.39 & 0.94 & \multirow{2}{*}{C2}\\
 & & & & & 0.95 & 0.37 & 7.76 & 5.64 & 1.38 & \\
\bottomrule
\end{tabular*}
\end{sidewaystable}

\begin{figure}[!htbp]
    \centering
    {\includegraphics[width=0.49\linewidth]{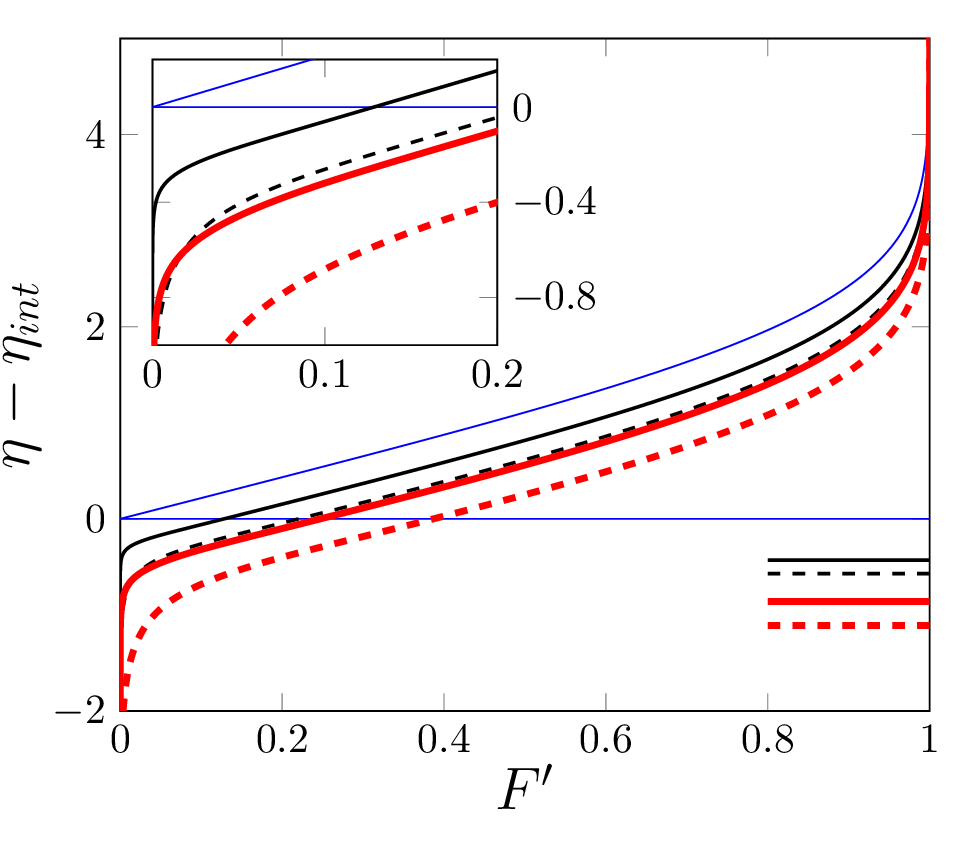}
    \includegraphics[width=0.49\linewidth]{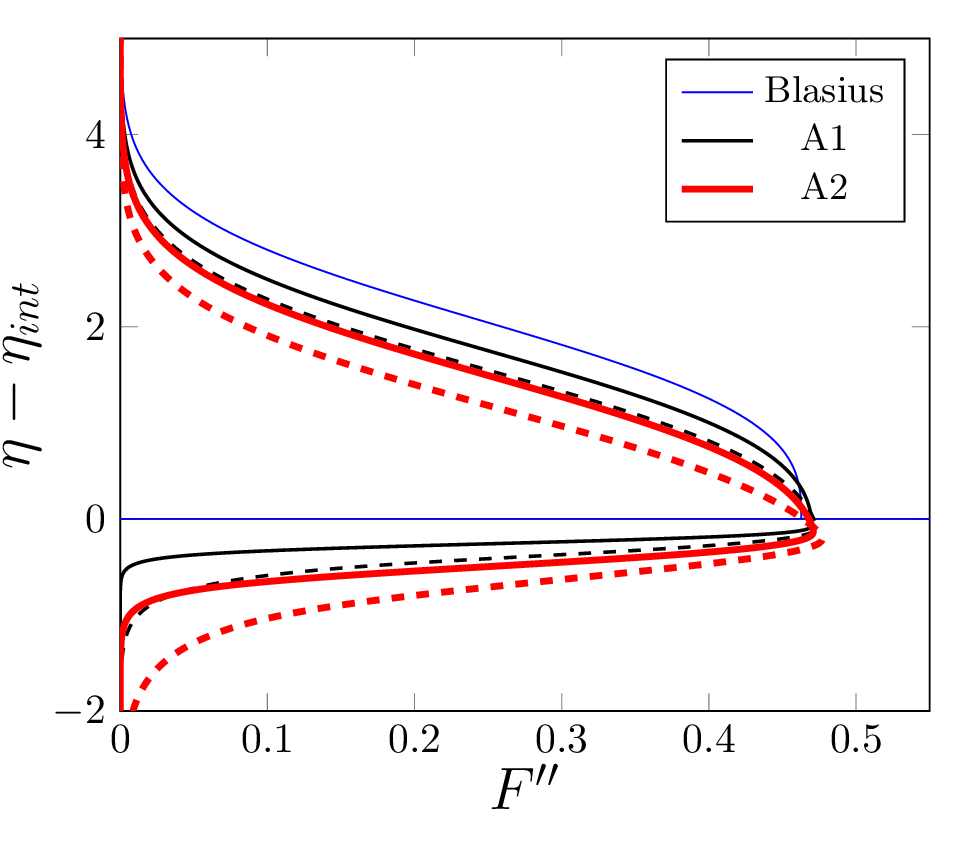}}
    \caption[Wall-normal profiles of $F^\prime$ (incompressible flow).]{First (left plot) and second (right plot) derivative of $F$ as a function of $\eta$ for an incompressible flow ($\Mach=0.01$). The black and red curves show cases A1 and A2, respectively, for $\theta_{fp}=0.85$ (solid) and $\theta_{fp}=0.95$ (dashed). The Blasius solution over a non-permeable wall is plotted in blue for comparison. The top boundary of the interfacial region is located at $\eta_{int}$ (blue line), while the bottom one is marked by the horizontal lines for each case.}
    \label{fig:results_mach001}
\end{figure}
\begin{figure}[ht!]
    \centering
    {\includegraphics[width=0.49\linewidth]{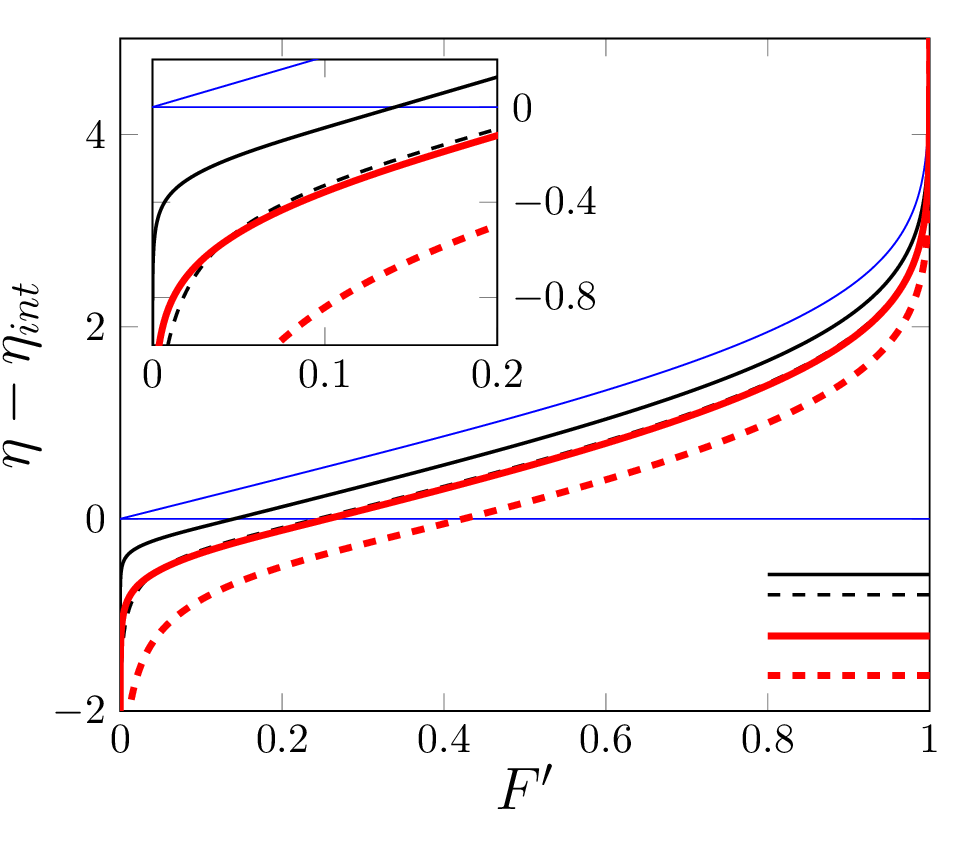}\includegraphics[width=0.49\linewidth]{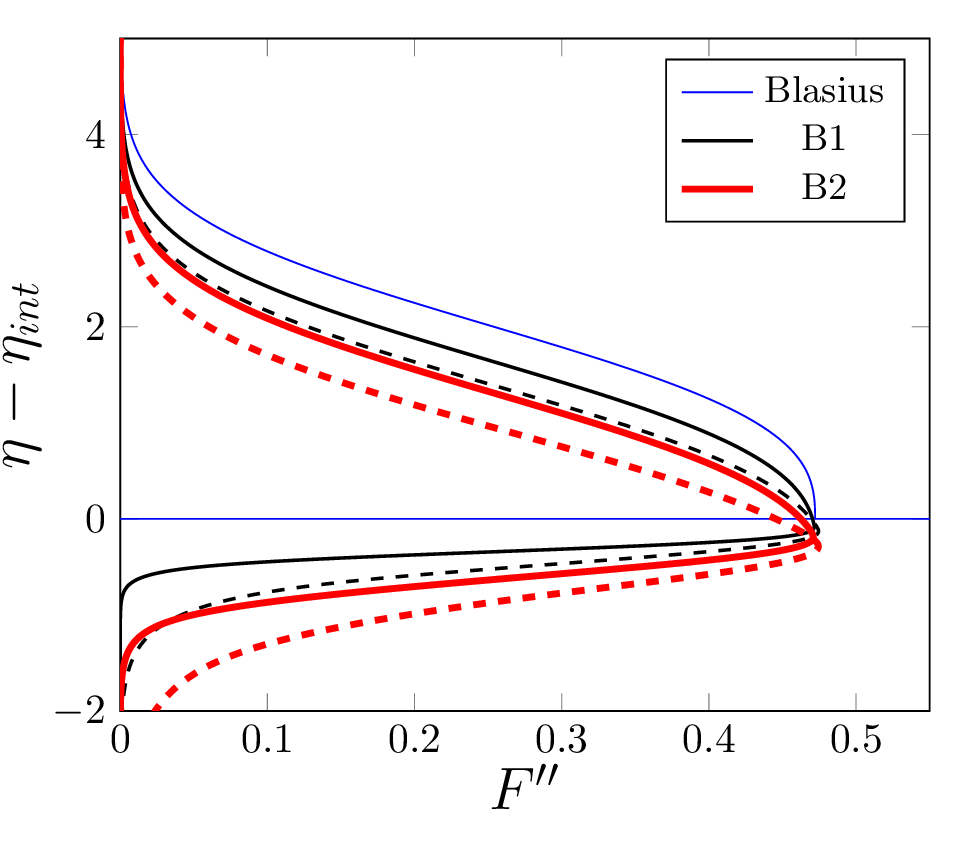}} \\
    {\includegraphics[width=0.49\linewidth]{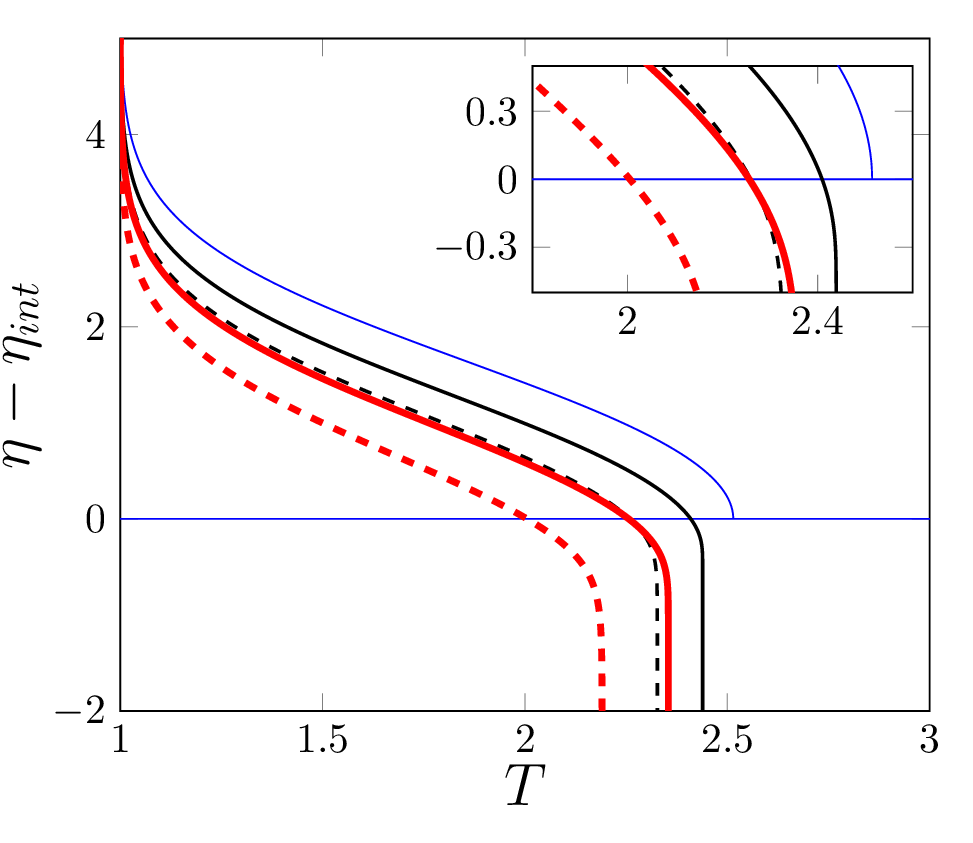}\includegraphics[width=0.49\linewidth]{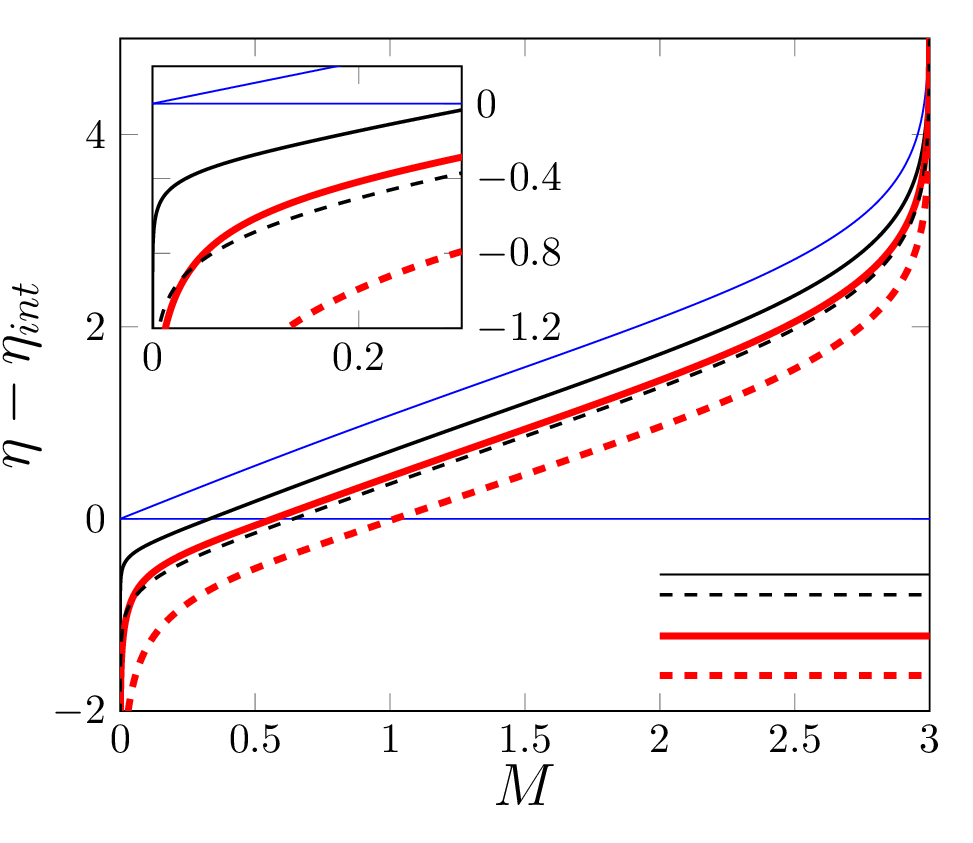}}
    \caption[Wall-normal profiles of $F^\prime$, $T$ and $M$ ($\Mach=3$).]{First (top left) and second derivative (top right) of $F$ as a function of $\eta$ for a supersonic flow at $\Mach=3$. The static temperature $T$ and the local Mach number $M$ are shown in the bottom-left and bottom-right plots, respectively. The black and red curves show cases B1 and B2, respectively, for $\theta_{fp}=0.85$ (solid) and $\theta_{fp}=0.95$ (dashed). The Blasius solution over a non-permeable wall is plotted in blue for comparison. The top boundary of the interfacial region is located at $\eta_{int}$ (blue line), while the bottom one is marked by the horizontal lines for each case.}
    \label{fig:results_mach3}
\end{figure}
\begin{figure}[ht!]
    \centering
    {\includegraphics[width=0.49\linewidth]{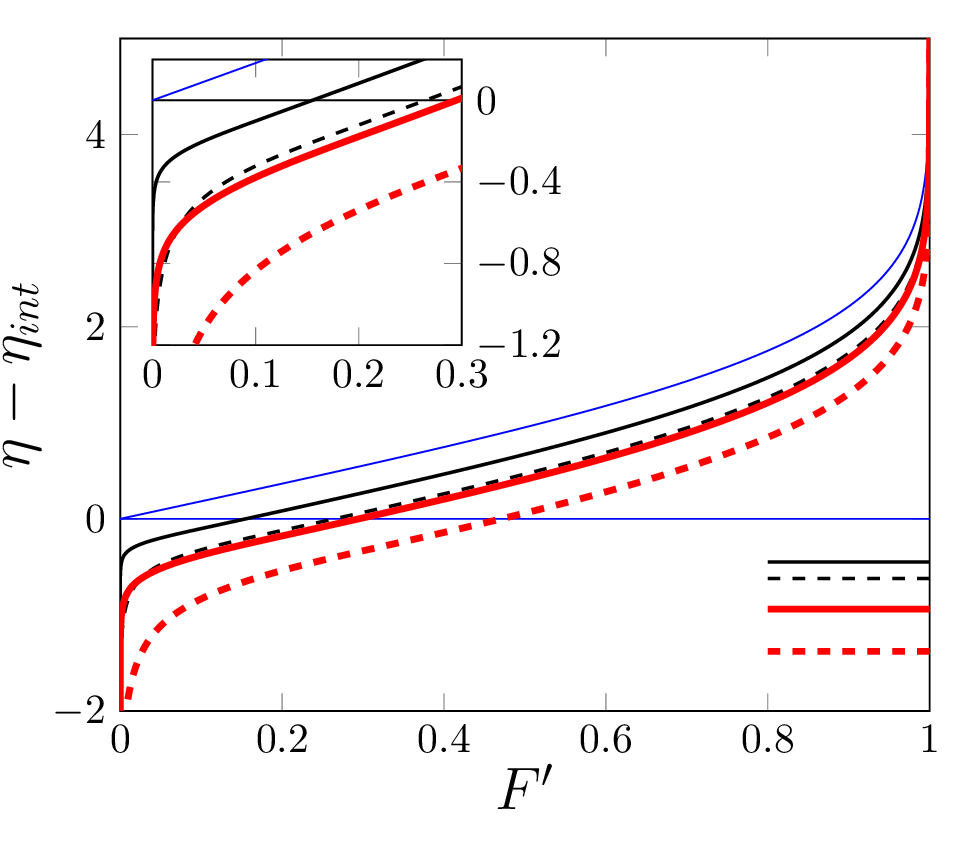}\includegraphics[width=0.49\linewidth]{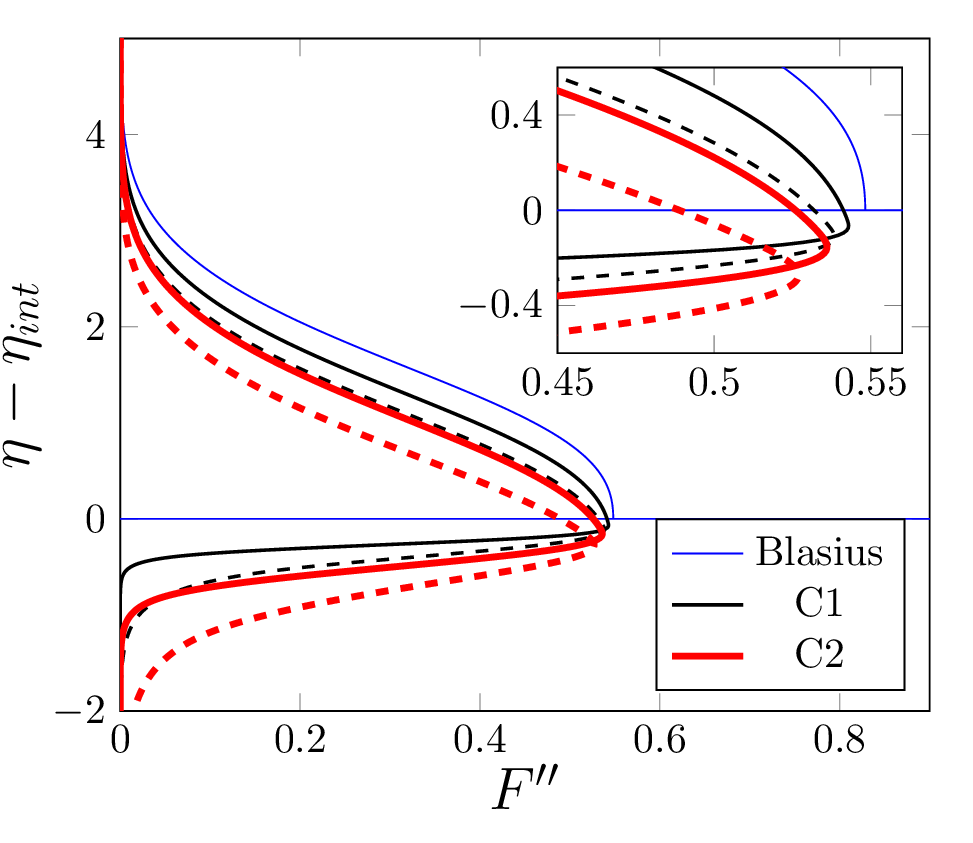}} \\
    {\includegraphics[width=0.49\linewidth]{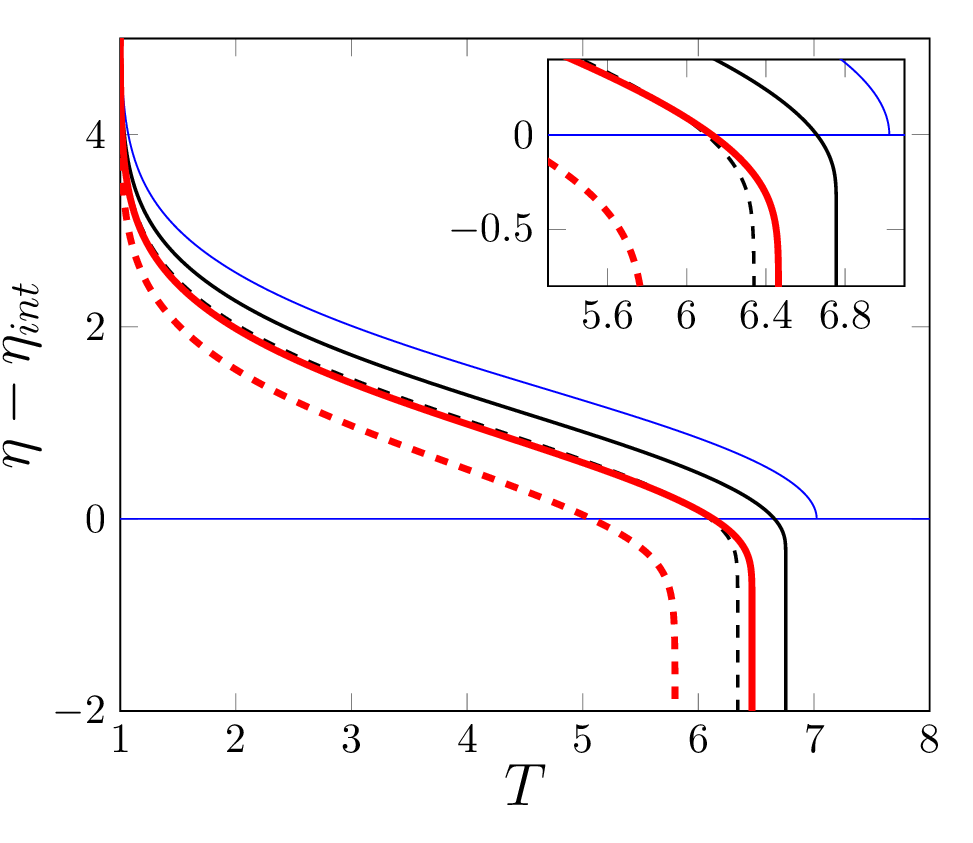}\includegraphics[width=0.49\linewidth]{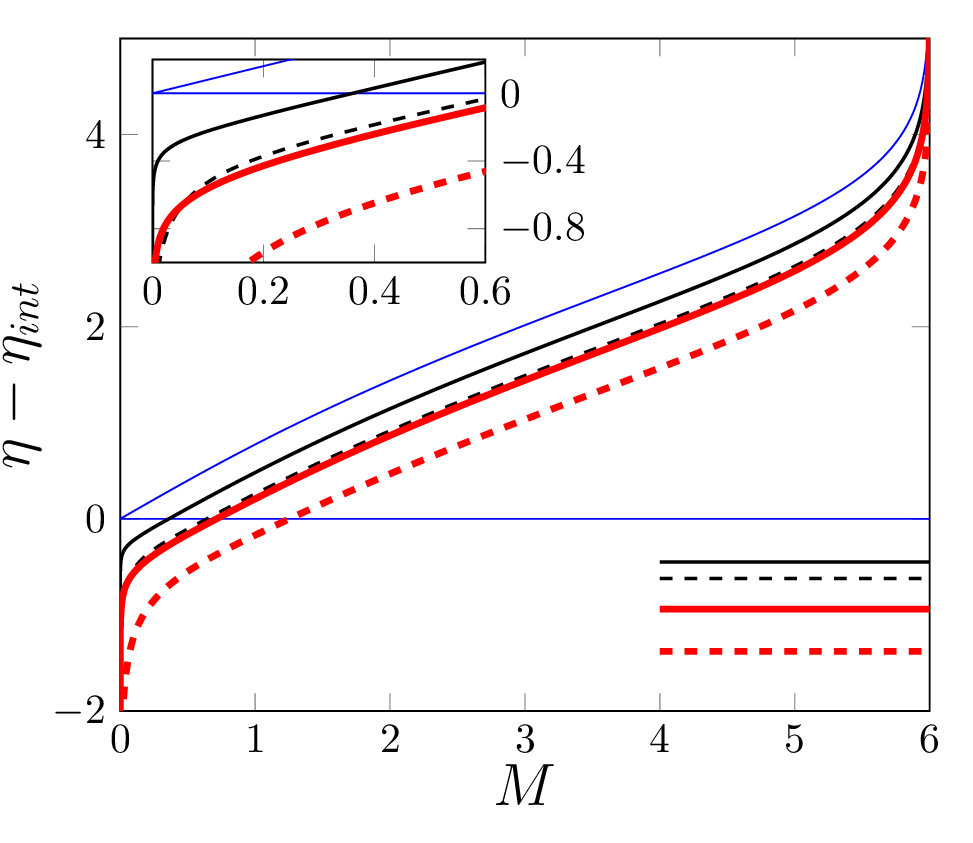}}
    \caption[Wall-normal profiles of $F^\prime$, $T$ and $M$ ($\Mach=6$).]{First (top left) and second derivative (top right) of $F$ as a function of $\eta$ for a supersonic flow at $\Mach=6$. The static temperature $T$ and the local Mach number $M$ are shown in the bottom-left and bottom-right plots, respectively. The black and red curves show cases C1 and C2, respectively, for $\theta_{fp}=0.85$ (solid) and $\theta_{fp}=0.95$ (dashed). The Blasius solution over a non-permeable wall is plotted in blue for comparison. The top boundary of the interfacial region is located at $\eta_{int}$ (blue line), while the bottom one is marked by the horizontal lines for each case.}
    \label{fig:results_mach6}
\end{figure}

The flow is studied for different values of the free-stream Mach number, the static pressure and temperature, the reference grain size $d_{g0}^\ast$ and the volume porosity below the interface $\theta_{fp}$. The combination of these physical parameters determines $\kappa_p$, $C_D$, $C_F$ and $(\theta_{\p ff})_p$. The values relevant to the investigated cases are listed in table \ref{tab:parameters}. For every case, the letter (A, B or C) defines a set of free-stream conditions, while the number (1 or 2) defines $d_{g0}^\ast$. The dimensional thickness of the Blasius boundary layer at $x^\ast = L^\ast$ is $\delta_{99}^\ast = \widehat{\delta}_{99} \left(\nu_\infty^\ast L^\ast/U_\infty^\ast\right)^{1/2}$. Here $\widehat{\delta}_{99} = \sqrt{2}\int_0^{\eta_{99}}T(\breve{\eta})\d\breve{\eta}$ increases with $\Mach$ and is a function of $\gamma$, $\Prandtl$ and the temperature boundary condition at the bottom solid wall. The values of $\delta_{99}^\ast$, estimated in table \ref{tab:parameters}, are in good agreement with wind tunnel measurements (refer to table \ref{tab:experimental_blayers_porous}). The stagnation pressure is $p_\circ^\ast$, the stagnation temperature is $T_\circ^\ast$, the dimensional wall temperature is $T_w^\ast$ and the adiabatic recovery temperature of the Blasius solution is $T_{ad,w}^\ast$. 

Results for an incompressible, isothermal ($T_{av}=1$) boundary layer at $\Mach=0.01$ (cases A1 and A2) are shown in figure \ref{fig:results_mach001}. The first and second derivative of the streamfunction $F$ are plotted as functions of $\eta$ for two different values of the volume porosity $\theta_{fp}$ and characteristic grain size $d_{g0}^\ast$ ($\kappa_p$). The top of the free fluid-porous interfacial region described in figure \ref{fig:interfacial_model} is located at $\eta=\eta_{int}$ (thin horizontal blue line in figure \ref{fig:results_mach001}) and its lower boundary is denoted by a thin horizontal black or red line for each case (also refer to figures/pdfs \ref{fig:results_mach3} and \ref{fig:results_mach6}). The Blasius solution for a flat-plate boundary layer is also drawn (blue curve), its solid wall being located at the top of the interfacial region for comparison \cite{Breugem_Boersma_2005}, where $\theta_f=1$ and the Darcy and Forchheimer terms become zero. The slip velocity, obtained by evaluating \eqref{eq:streamwise_velocity} at the interface, increases with $\theta_{fp}$ and $d_{g0}^\ast$, in qualitative agreement with the results of \cite{Tsiberkin_2018a,Tsiberkin_2018b}, which showed increasing interfacial velocity for increasing values of the control parameter $\kappa_p^2$ without the Forchheimer correction. The incompressible intrinsic shear stresses $\tau = (F^{\prime\prime} - F^\prime\theta_f^\prime/\theta_f)/\theta_f$ (i.e. the shear stresses multiplied by $(2x)^{1/2}$, refer to \cite[equations 116-117]{Breugem_Boersma_Uittenbogaard_2005}) are mostly influenced by $F^{\prime\prime}$ and grow to a peak located at the interface in all cases. Both $F^\prime$ and $F^{\prime\prime}$ undergo a rapid decay across the interface and within the porous region. 

The results of the supersonic cases are shown in figure \ref{fig:results_mach3} and \ref{fig:results_mach6}. Adiabatic boundary conditions were imposed at the solid wall below the substrate. The profiles for $F^\prime$ and $F^{\prime\prime}$ are shown along with the temperature $T$ (bottom left) and the local Mach number $M=u^\ast/(\gamma\mathcal{R}^\ast T^\ast)^{1/2}=\Mach\,F^\prime/(\theta_f T^{1/2})$ (bottom right). The slip velocity increases and $F^{\prime\prime}$ decreases with increasing $d_{g0}^\ast$ and $\theta_{fp}$ for constant free-stream conditions. Moreover, the temperature $T$ does not recover the adiabatic value of the compressible Blasius solution and it reduces sharply at the interface when a porous substrate is introduced. The reduction becomes more marked with increasing $\kappa_p$ and $\theta_{fp}$. The peak of $F^{\prime\prime}$ is not sensitive to the geometry of the substrate. The influence of the porous substrate on the velocity profiles extends far from the interface, where $F^{\prime\prime}$ is reduced. As the $F^\prime$ profile shifts towards the interface and the magnitude of $T$ decreases, the local Mach number $M$ increases across the boundary layer (refer to the bottom right plots of figures/pdfs \ref{fig:results_mach3} and \ref{fig:results_mach6}). The insets in the bottom-right plot of figures/pdfs \ref{fig:results_mach3} and \ref{fig:results_mach6} show that $M$ is always well below unity near the bottom boundary of the interfacial region where the top solid cubes are located. This finding rules out the presence of shock waves and choking.

\begin{figure}[ht!]
    \centering
    {\includegraphics[width=0.49\linewidth]{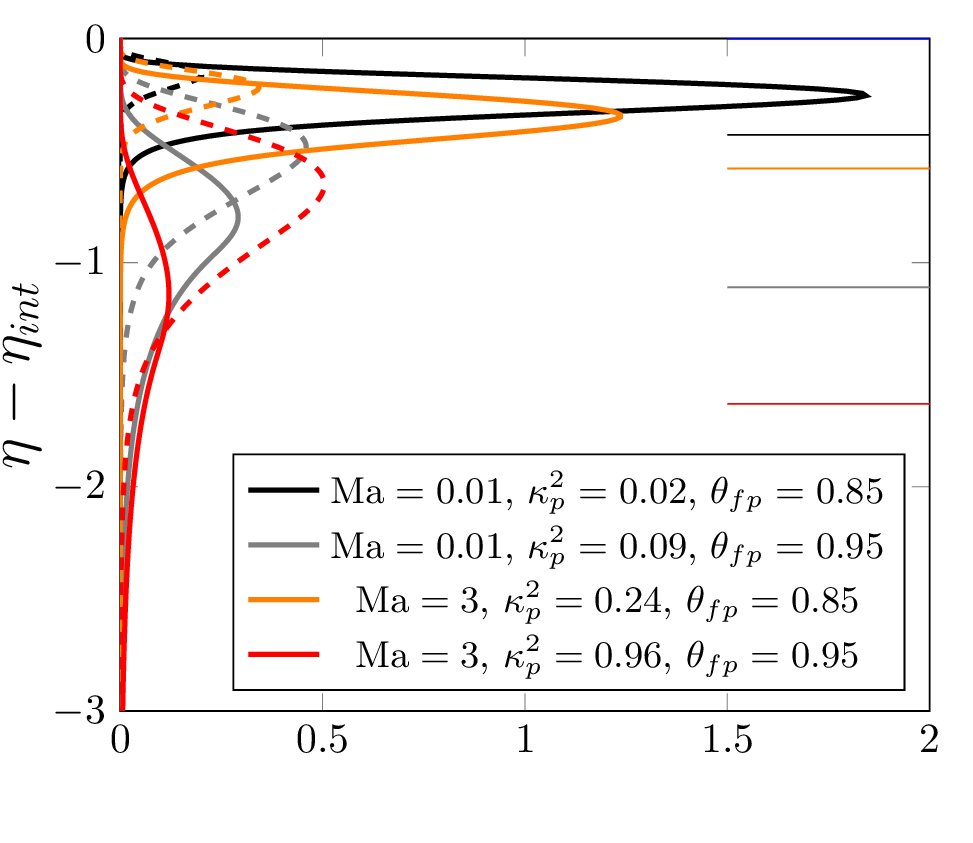}
    \includegraphics[width=0.49\linewidth]{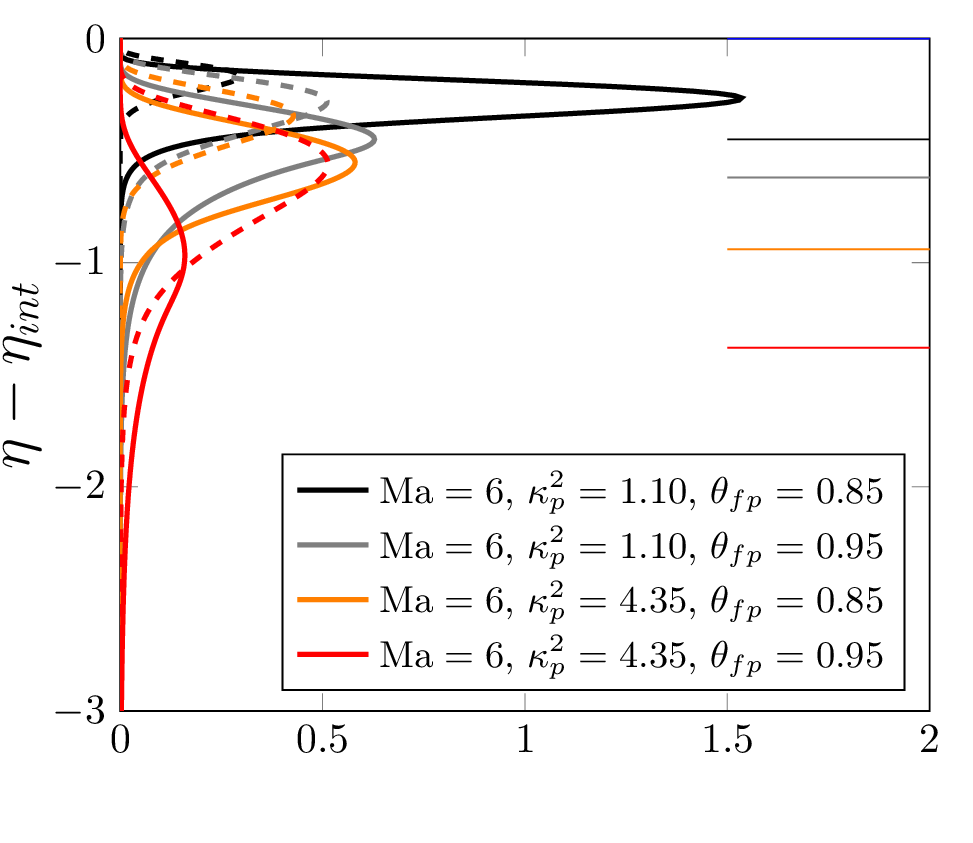}}
    \caption[Distribution of the Darcy and Forchheimer terms across the interface.]{Wall-normal profiles of the Darcy term (solid curves) and Forchheimer term (dashed curves) within the interfacial region. Left plot: cases A1, A2, B1 and B2. Right plot: C1 and C2.}
    \label{fig:darcy-forchheimer}
\end{figure}

Figure \ref{fig:darcy-forchheimer} shows that the Darcy term and Forchheimer term in the momentum equation \eqref{eq:self-similar_porous_momentum} reach a comparable magnitude for all values of $\Mach$, $\kappa_p$ and $\theta_{fp}$. The Forchheimer correction is considerably lower than the Darcy drag for low $\kappa_p$ (low $d_{g0}^\ast$) and $\theta_{fp}$, but increases and becomes slightly larger as either $\kappa_p$ or $\theta_{fp}$ increase. This behavior is reported for all Mach numbers as shown in the right plot of figure \ref{fig:darcy-forchheimer}, in which the free-stream conditions are fixed and the grain size and the porosity are gradually increased. Because the coefficient $C_F$ is assumed to be a function of $d_{g0}^\ast$ instead of $d_{g}^\ast$, this result suggests that the local similarity assumption holds better at moderate $\theta_{fp}$ and $\kappa_p$ \cite{Tsiberkin_2018b} for all Mach numbers.

\begin{figure}[ht!]
    \centering
    {\includegraphics[width=0.49\linewidth]{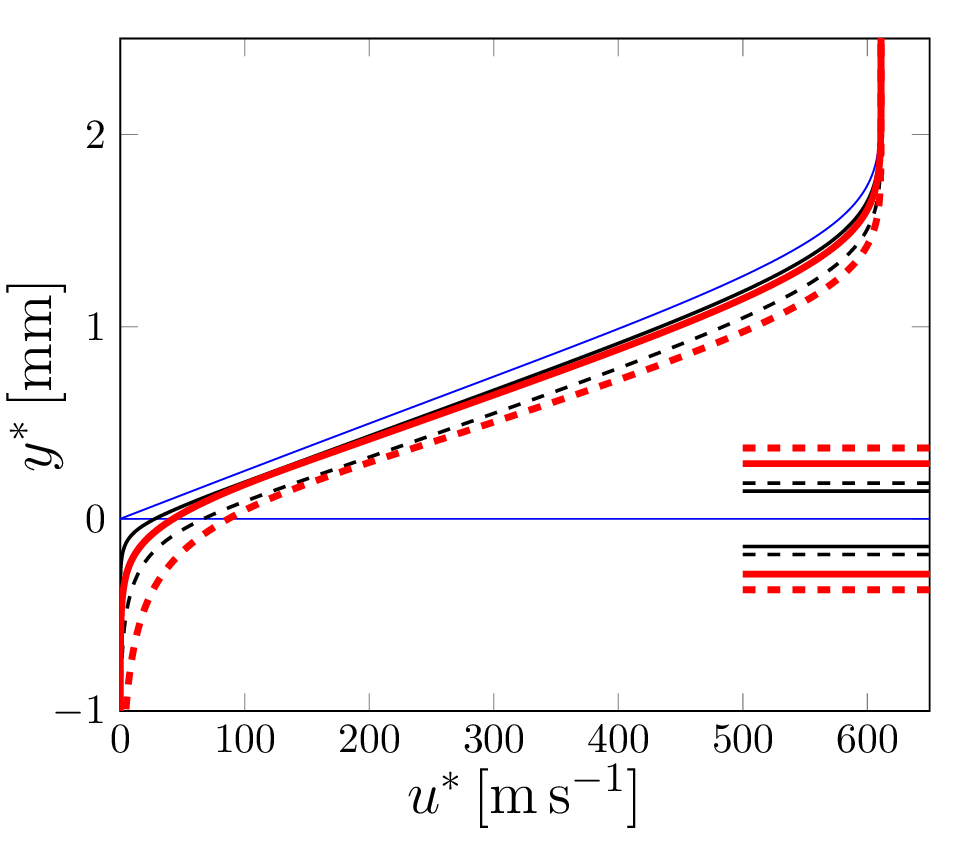}
    \includegraphics[width=0.49\linewidth]{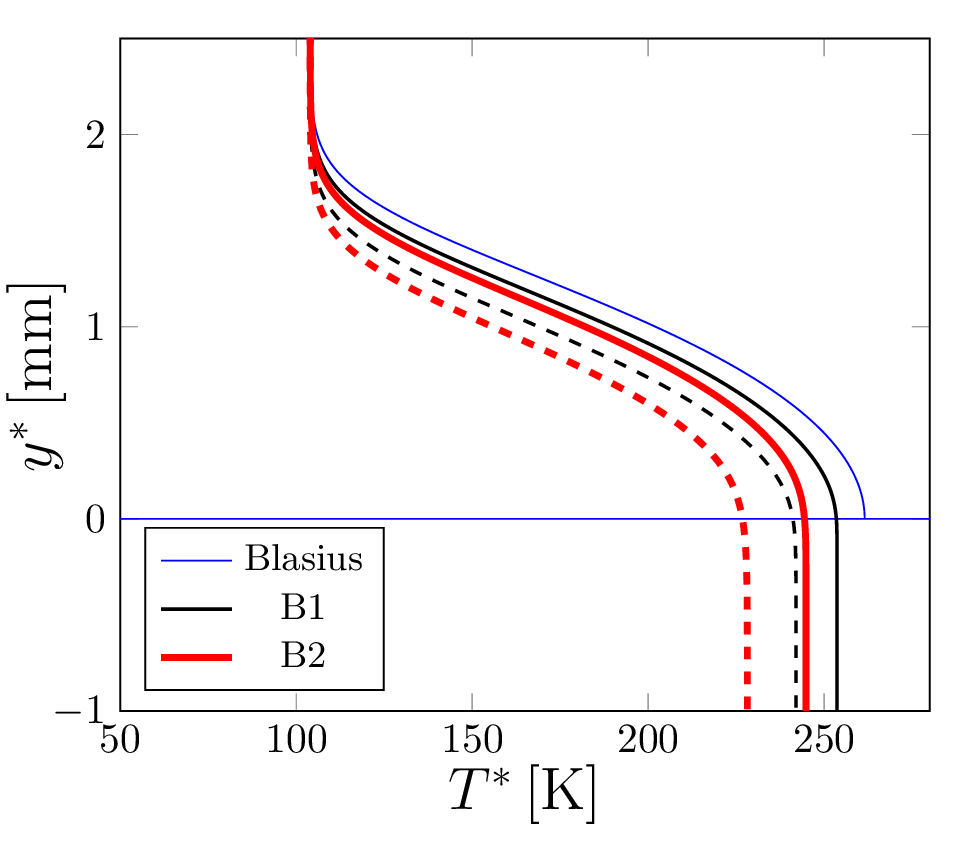}}
    {\includegraphics[width=0.49\linewidth]{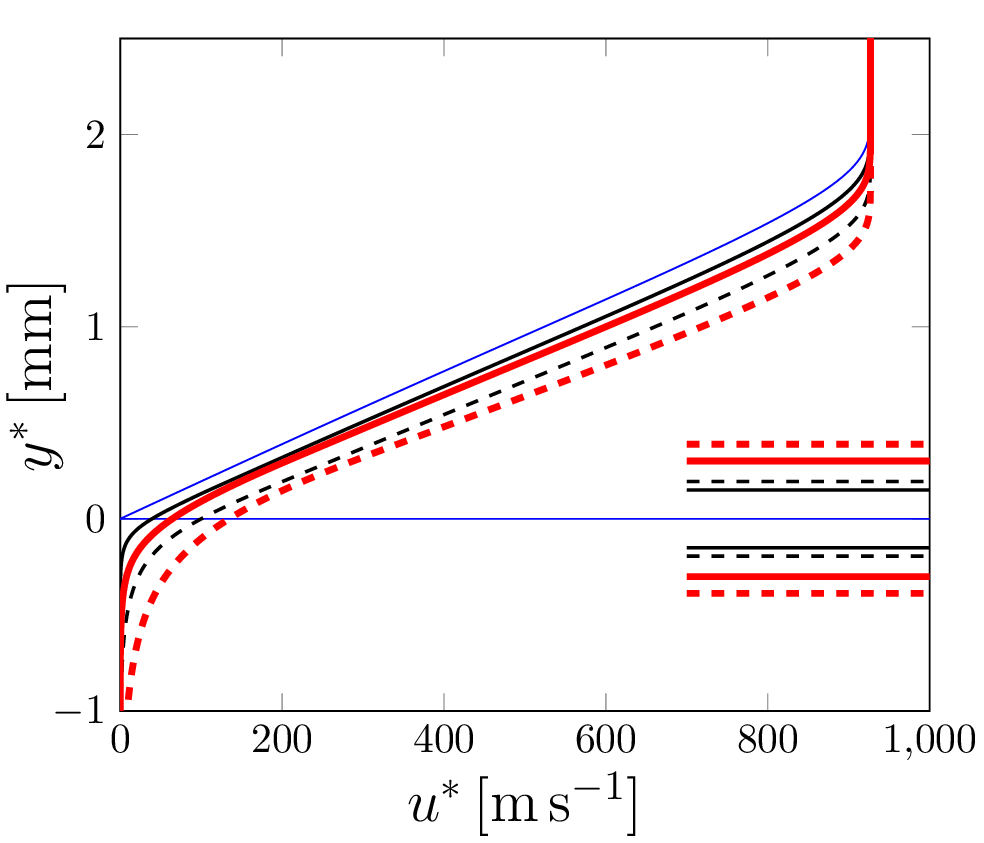}
    \includegraphics[width=0.49\linewidth]{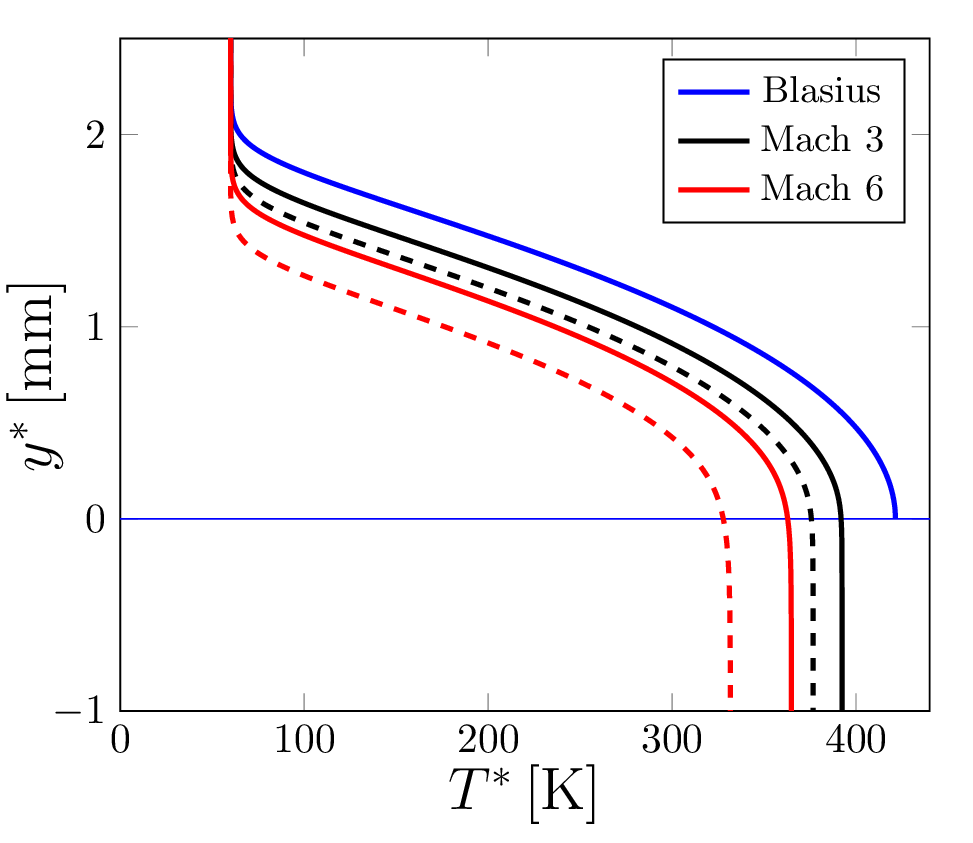}}
    \caption[Dimensional velocity and temperature profiles of a boundary layer over a porous substrate.]{Dimensional velocity and temperature profiles at a free-stream Mach number $3$ (cases C1 and C2, top) and $6$ (cases C1 and C2, bottom). The interfacial thickness is $\delta_{int}^\ast=408\,\si{\upmu\meter}$ (B1 and C1, $\theta_{fp}=0.85$), $525\,\si{\upmu\meter}$ (B1 and C1, $\theta_{fp}=0.95$), $815\,\si{\upmu\meter}$ (B2 and C2, $\theta_{fp}=0.85$), $1051\,\si{\upmu\meter}$ (B2 and C2, $\theta_{fp}=0.95$). The effect of varying $\theta_{fp}$ and $d_{g0}^\ast$ is shown. The Mach-3 and Mach-6 adiabatic Blasius solution is plotted in blue.}
    \label{fig:dim_profiles}
\end{figure}

The dimensional profiles of the streamwise velocity $u^\ast$ and the static temperature $T^\ast$ are computed at a distance $x^\ast=L^\ast$ from the leading edge and are plotted in figure \ref{fig:dim_profiles} for the Mach-3 cases B1 and B2 and the Mach-6 cases C1 and C2. The physical coordinate $y^\ast$ is computed by $y^\ast = x^\ast(2/\Reynolds_x)^{1/2} \int_0^\eta T\left(\breve{\eta}\right)\d\breve{\eta}$, where $\Reynolds_x$ is the Reynolds number based on $x^\ast$ \cite{Stewartson_1964}. The boundary-layer thickness of the Blasius solution (blue curve) compares well to the measurements of \cite{Graziosi_Brown_2002} and \cite{Maslov_Shiplyuk_Sidorenko_Arnal_2001} (refer to table \ref{tab:experimental_blayers_porous}). When the substrate is sufficiently thick, the velocity profile decays quasi-exponentially because only the diffusion and linear Darcy terms remain dominant underneath the interface and the momentum equation \eqref{eq:self-similar_porous_momentum} reduces to the Darcy-Brinkman equation
\begin{equation}
    \left(\frac{\mu}{T}F^{\prime\prime}\right)^\prime - C_D\mu T\frac{\left(1-\theta_f\right)^2}{\theta_f^2}F^\prime = 0
\end{equation}
This behavior is shown in figure \ref{fig:darcy-forchheimer}: the Darcy terms (solid curves) decay more readily than the Forchheimer terms (dashed curves) underneath the interface. Hence, if the substrate is sufficiently thick, both $F^{\prime}$ and $F^{\prime\prime}$ are small at the bottom solid wall ($F^{\prime\prime}\cong 10^{-4}$), which makes the choice of $\eta_{int}$ arbitrary. Because the depth of the porous substrate is kept constant at $\eta_{int}=10$ in the $\eta$-space and varies slightly in the $y$-space, the physical plots in figure \ref{fig:dim_profiles} are offset and centered at the interface. Again, the velocity profiles shift towards the interface and the boundary-layer thickness is reduced as both $d_{g0}^\ast$ and $\theta_{fp}$ increase, while the temperature at the interface decreases. To better compare the porous-plate and the solid-plate results, the center of the interfacial region and the wall of the non-porous solution are placed at the same height in figure \ref{fig:dim_profiles}. The upper and lower boundaries of the interface are here denoted by the horizontal black and red lines. The departure of the velocity profile from the compressible Blasius solution is very small in the case C1, $\theta_{fp}=0.85$. Only for larger porosities and grain sizes the seepage becomes significant. This finding agrees qualitatively with the recent experiments of \cite{Running_Bemis_Hill_Borg_Redmond_Jantze_Scalo_2023}, who studied the development of a boundary layer over a porous substrate. Their measurements agreed well with the numerical results obtained from the computation of a no-slip, flat-plate, boundary-layer flow. They used a silicon-carbide foam porous insert with $\theta_{fp}=0.862$ and 3.9 pores per linear millimeter, which corresponds to $d_{g0}^\ast=130\,\si{\upmu\meter}$ in the present model. The present results suggest that porous media with higher $\theta_{fp}$ and $d_{g0}^\ast$ should be used to affect the base flow significantly.

\begin{figure}[ht!]
    \centering
    {\includegraphics[width=0.49\linewidth]{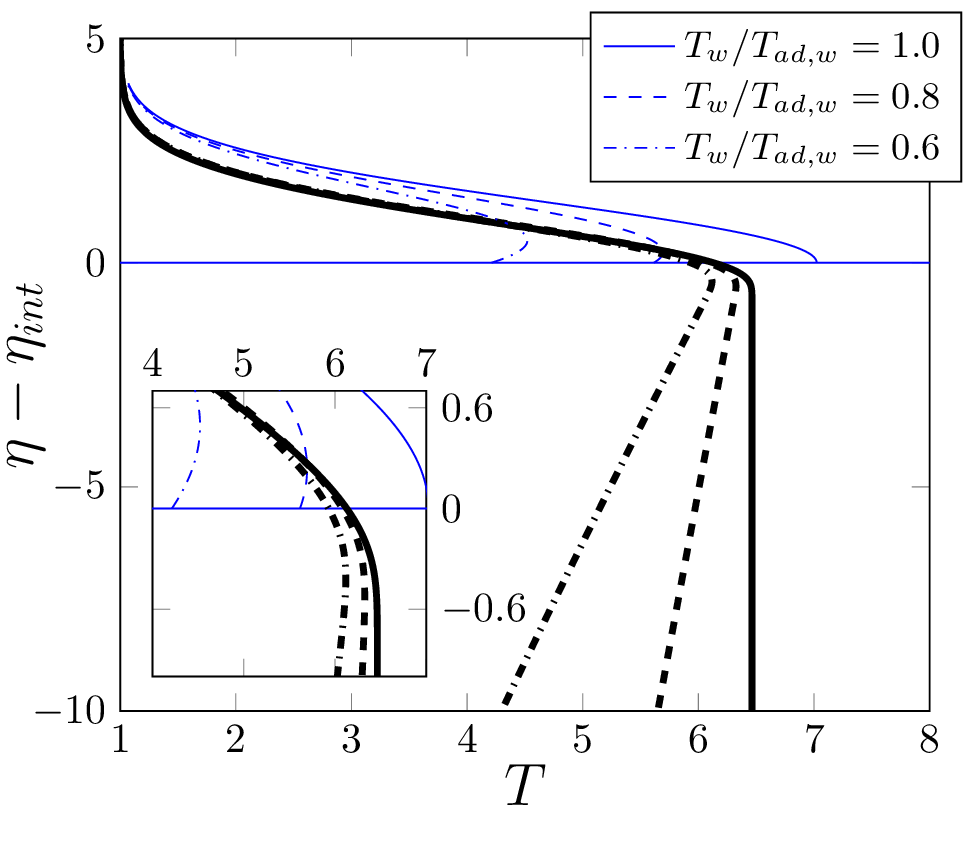}
    \includegraphics[width=0.49\linewidth]{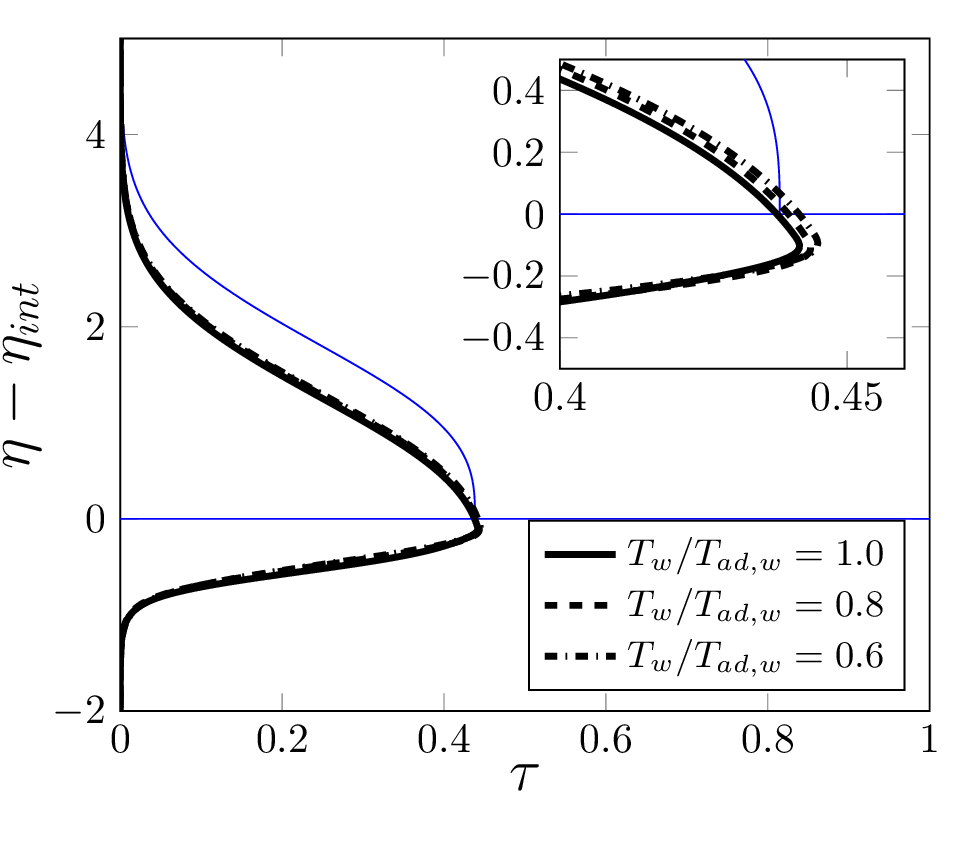}}
    \caption[Effect of the temperature boundary conditions at the wall at $\Mach=6$.]{Effect of the temperature boundary conditions at the bottom solid wall on the temperature profiles (left plot) and the shear stresses (right plot) at $\Mach=6$. The curves are computed for different values of $T_w/T_{ad,w}$, where $T_{ad,w}=7.02$ is the adiabatic wall temperature of the Blasius solution. The blue curves represent the Blasius solution and the black ones the porous substrate solution (C2, $\theta_{fp}=0.85$). The adiabatic recovery temperature at the solid wall is $T_{ad,w}=6.46$.}
    \label{fig:temperature_BC}
\end{figure}

\begin{figure}[ht!]
    \centering
    {\includegraphics[width=0.49\linewidth]{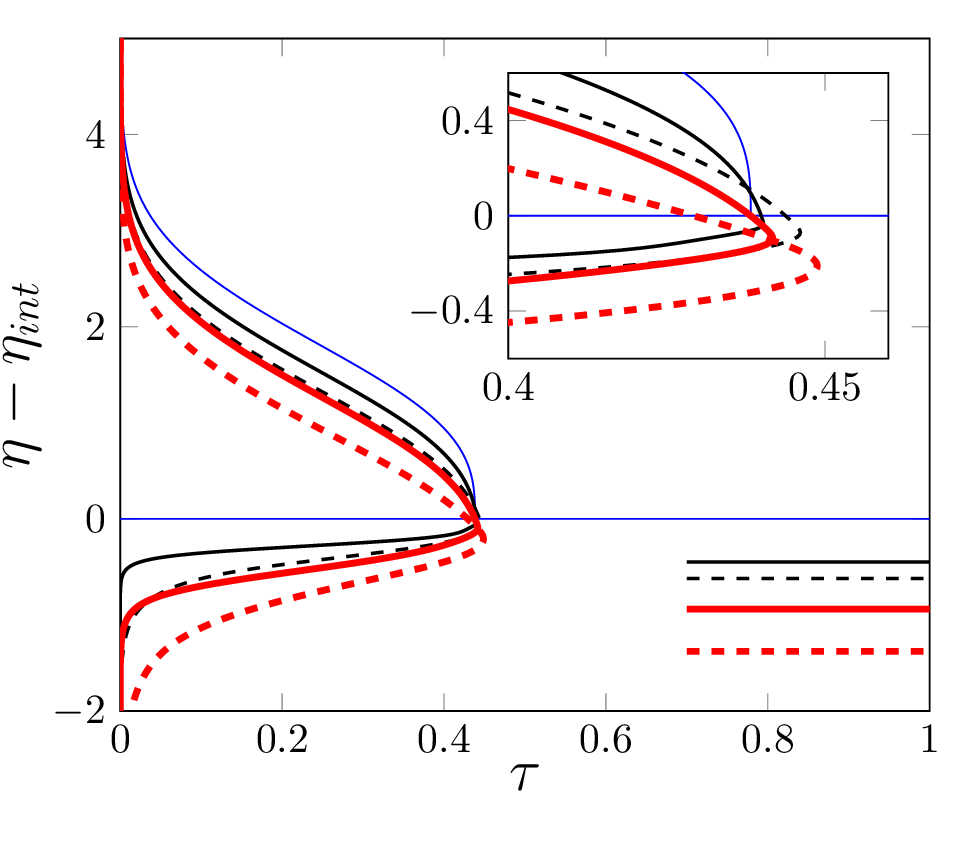}
    \includegraphics[width=0.49\linewidth]{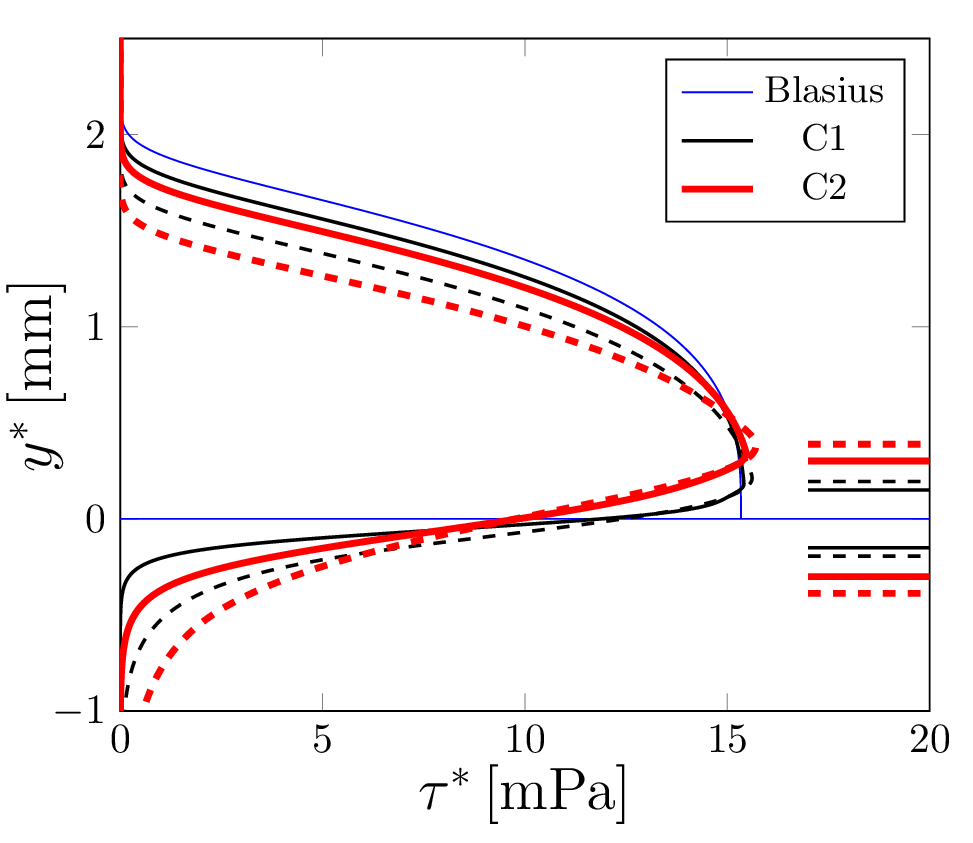}}
    \caption[Distribution of the intrinsic shear stresses.]{Distribution of the intrinsic shear stresses for variable geometry ($\kappa_p$ and $\theta_{fp}$) in transformed (left) and real, dimensional coordinates at $x^\ast=L^\ast$ (right).}
    \label{fig:shear_stresses}
\end{figure}

The effect of varying the temperature of the bottom solid wall is shown in figure \ref{fig:temperature_BC} for the case C2, $\theta_{fp}=0.85$. Because natural convection within the medium is not captured by the present model, wall-cooling conditions for which the bottom-wall temperature $T_w$ is lower than the adiabatic recovery temperature of the substrate are considered. The temperature curves (figure \ref{fig:temperature_BC}, left plot) approach the adiabatic case as the ratio $T_w/T_{ad,w}$ increases. The temperature at the interface decreases from $T_w/T_{ad,w}=1$ to $T_w/T_{ad,w}=0.85$ with respect to the Blasius solution. However, as the ratio decreases further, the temperature at the interface remains higher and decreases linearly where the flow is stagnant. Equations \eqref{eq:coupled_por_BL_enthalpy} and \eqref{eq:self-similar_porous_enthalpy} then reduce to the steady homogeneous heat conduction equation as in the model of \cite{Nield_Kutsenov_2003}. Unlike in the Blasius boundary layer, where the temperature boundary condition at the solid wall directly affects the value of $F^{\prime\prime}$, the temperature condition imposed at the bottom of the porous substrate has no influence on $F^{\prime\prime}$ at the interface with the fluid region. The $F^\prime$ and $F^{\prime\prime}$ profiles are unaffected by the ratio $T_w/T_{ad,w}$ (refer to the right plot of figure \ref{fig:temperature_BC}). The results in figure \ref{fig:temperature_BC} stem from the assumption of local thermal equilibrium of the fluid and solid phases, which may not be valid if the temperature at the bottom solid wall $T_w$ departs the adiabatic condition $T_{w,ad}$ significantly. The distribution of the compressible intrinsic shear stresses $\tau = \mu ( F^{\prime\prime} - F^\prime \theta_f^\prime/\theta_f)/(\theta_f T)$, shown in the plots in figure \ref{fig:shear_stresses}, is mostly affected by the geometry of the substrate, whereas the effect of $T_w/T_{ad,w}$ is again negligible. The peaks increase slightly and appear to be always located near the top boundary of the interface. A sharp reduction in the intrinsic shear stresses occurs in the free-fluid region above the interface when a substrate of high porosity is introduced.

%%%%%%%%%%%%%%%%%%%%%%%%%%%%%%%%%%%%%%%%%%%%%%%%%%%%
%%%%%%%%%%%%%%%%%%%%%%%%%%%%%%%%%%%%%%%%%%%%%%%%%%%%
%%%%%%%%%%%%%%%%%%%%%%%%%%%%%%%%%%%%%%%%%%%%%%%%%%%%
%%%%%%%%%%%%%%%%%%%%%%%%%%%%%%%%%%%%%%%%%%%%%%%%%%%%
%%%%%%%%%%%%%%%%%%%%%%%%%%%%%%%%%%%%%%%%%%%%%%%%%%%%

\FloatBarrier
\section{Summary}
For the first time, a self-similar compressible laminar boundary layer flowing over an isotropic porous substrate of streamwise-increasing permeability is studied by asymptotic and numerical methods. This setup was considered by \cite{Tsiberkin_2016} for the incompressible case. Porous substrates with variable permeability may soon be manufactured and their mathematical description allows for a self-similar solution of the boundary-layer equations. The solution includes a linear Darcy term and a quadratic Forchheimer correction. The volume averaged momentum and enthalpy balance equations become parabolic in the limit of high Reynolds and small Darcy numbers. The effect of the porous substrate is distilled in the distributions of the volume and surface porosity, the control parameter $\kappa_p$ and the Forchheimer coefficient $C_F$. The thicknesses of the interface and the boundary layer are comparable and the volume and surface porosity vary smoothly therein. The wall-normal profiles of the streamwise velocity $F^\prime$, its wall-normal derivative $F^{\prime\prime}$ and the temperature $T$ have been computed for different values of the Mach number, the control parameter and the volume porosity below the interface. The profiles bear a strong resemblance to the Blasius solution for low values of $\kappa_p$ and $\theta_{fp}\leq 0.85$. The slip velocity at the porous-free fluid interface increases and the shear stresses decrease sharply, as the volume porosity approaches unity. Wall cooling at the bottom solid boundary affects neither the velocity profiles nor the shear stresses when the porous substrate is sufficiently thick. A sharp reduction in the shear stresses and the static temperature is observed above the interface. This result shows that the introduction of a porous substrate of high porosity and permeability can substantially alter the properties of a supersonic laminar boundary layer, which is significant to flow control applications \cite{Maslov_Mironov_Poplavskaya_Kirilovskiy_2019,Running_Bemis_Hill_Borg_Redmond_Jantze_Scalo_2023}. 

%%%%%%%%%%%%%%%%%%%%%%%%%%%%%%%%%%%%%%%%%%%%%%%%%%%%%%%%%%%%%%%%%%%%
%%%%%%%%%%%%%%%%%%%%%%%%%%%%%%%%%%%%%%%%%%%%%%%%%%%%%%%%%%%%%%%%%%%%
%%%%%%%%%%%%%%%%%%%%%%%%%%%%%%%%%%%%%%%%%%%%%%%%%%%%%%%%%%%%%%%%%%%%
%%%%%%%%%%%%%%%%%%%%%%%%%%%%%%%%%%%%%%%%%%%%%%%%%%%%%%%%%%%%%%%%%%%%
%%%%%%%%%%%%%%%%%%%%%%%%%%%%%%%%%%%%%%%%%%%%%%%%%%%%%%%%%%%%%%%%%%%%

\section*{Statements and Declarations}

\noindent
\textbf{Funding} The authors acknowledge the support of the US Air Force through the AFOSR grant FA8655-21-1-7008 (International Program Office - Dr Douglas Smith). PR has also been supported by EPSRC (Grant No. EP/T01167X/1). 

\noindent
\textbf{Competing interests} The authors have no relevant financial or non-financial interests to disclose.

\noindent
\textbf{Data availability statement} The datasets generated during and/or analysed during the current study are available from the corresponding author on reasonable request.

\noindent
\textbf{Contributions} LF wrote and proof-read this manuscript, developed, implemented and visualised the presented material. PR conceptualised and supervised this work and contributed to writing of the manuscript. 

%%%%%%%%%%%%%%%%%%%%%%%%%%%%%%%%%%%%%%%%%%%%%%%%%%%%%%%%%%%%%%%%%%%%
%%%%%%%%%%%%%%%%%%%%%%%%%%%%%%%%%%%%%%%%%%%%%%%%%%%%%%%%%%%%%%%%%%%%
%%%%%%%%%%%%%%%%%%%%%%%%%%%%%%%%%%%%%%%%%%%%%%%%%%%%%%%%%%%%%%%%%%%%
%%%%%%%%%%%%%%%%%%%%%%%%%%%%%%%%%%%%%%%%%%%%%%%%%%%%%%%%%%%%%%%%%%%%
%%%%%%%%%%%%%%%%%%%%%%%%%%%%%%%%%%%%%%%%%%%%%%%%%%%%%%%%%%%%%%%%%%%%

\appendix
\section{The Forchheimer term in the momentum equation}\label{app:forchheimer}
The self-similar momentum equation \eqref{eq:self-similar_porous_momentum} is derived from the dimensional streamwise momentum balance equation \eqref{eq:momentum_balance_dim_porous}, where only the leading-order terms are retained and the volume-averaging operators are omitted,
\begin{multline}\label{eq:momentum_balance_dim_porous_appendix}
    \theta_f\rho^\ast U^\ast \frac{\p U^\ast}{\p x^\ast} + \theta_f \rho^\ast V^\ast \frac{\p U^\ast}{\p y^\ast} = \frac{\p}{\p y^\ast}\left[\mu^\ast\frac{\p\left(\theta_f U^\ast\right)}{\p y^\ast}\right] + \\
    - \theta_f^2\frac{\mu^\ast U^\ast}{K^\ast} - \theta_f^2 \frac{c_F^\ast}{K^\ast}\rho^\ast U^{\ast2} + \order{\Reynolds^{-1}},
\end{multline}
and the form of $K^\ast\left(d_g^\ast\right)$ and $c_F^\ast\left(d_g^\ast=d_{g0}^\ast\right)$ that allows for a self-similar solution is used
\begin{subequations}
\begin{align}
    K^\ast\left(d_g^\ast\right) &= \frac{\theta_f^3}{\left(1-\theta_f\right)^2} \frac{d_g^{\ast2}}{A}, \\
    c_F^\ast\left(d_{g0}^\ast\right) &= \frac{\theta_f}{1-\theta_f} \frac{d_{g0}^{\ast}}{B}.
\end{align}
\end{subequations}

The terms in \eqref{eq:momentum_balance_dim_porous_appendix} are scaled by the reference quantities
\begin{multline}
    \frac{\rho_\infty^\ast U_\infty^\ast}{L^\ast} \left(\theta_f\rho U\frac{\p U}{\p x} + \theta_f\rho V \frac{\p U}{\p y}\right) = \frac{\mu_\infty^\ast U_\infty^\ast}{L^{\ast2}} \frac{\p}{\p y}\left[\mu \frac{\p\left(\theta_f U\right)}{\p y}\right] + \\
    - \frac{\mu_\infty^\ast U_\infty^\ast}{d_{g0}^{\ast2}} \left(\frac{d_{g0}^\ast}{d_g^\ast}\right)^2 \theta_f^2 \frac{A\left(1-\theta_f\right)^2}{\theta_f^3} \mu U + \\
    - \frac{\rho_\infty^\ast U_\infty^{\ast2}}{d_{g0}^{\ast2}} \left(\frac{d_{g0}^\ast}{d_g^\ast}\right)^2 \theta_f^2 \frac{A\left(1-\theta_f\right)^2}{\theta_f^3} \frac{\theta_f}{1-\theta_f} \frac{d_{g0}^\ast}{B} \rho U^2,
\end{multline}
and rearranged to obtain the leading-order balance for $A=\order{1}$ and $B=O(\Darcy^{-1/2})$ 
\begin{multline}
    \theta_f\rho U\frac{\p U}{\p x} + \theta_f\rho V \frac{\p U}{\p y} = \frac{\p}{\p y}\left[\mu \frac{\p \left(\theta_f u\right)}{\p y}\right] + \\
    - \frac{\theta_f^2}{\Reynolds\Darcy} \left(\frac{d_{g0}^\ast}{d_g^\ast}\right)^2 \left[A \frac{\left(1-\theta_f\right)^2}{\theta_f^3} \mu U + A \Reynolds\Darcy \frac{1-\theta_f}{\theta_f^2} \frac{\rho U^2}{B\Darcy^{1/2}}\right].
\end{multline}
By introducing the identities $\rho T=1$, $C_D = A/\kappa_p^2 = A (\Reynolds\Darcy)^{-1}$ and $C_F = A/(B\Darcy^{1/2})$, and the velocity components \eqref{eq:porous_base_flow_velocity_components}, the momentum balance becomes
\begin{multline}
    \left(\alpha x\right)^{a+b-1}\alpha\frac{\left(b-a\right)}{\theta_f}\left(\frac{\p F}{\p\eta}\right)^2 + \left(\alpha x\right)^{a+b}\frac{\p F}{\p\eta}\frac{\p}{\p x}\left(\frac{1}{\theta_f}\frac{\p F}{\p\eta}\right)_\eta + \\
    - \left(\alpha x\right)^{a+b}\left.\frac{\p F}{\p x}\right\vert_\eta \frac{\p}{\p\eta}\left(\frac{1}{\theta_f}\frac{\p F}{\p\eta}\right) - \left(\alpha x\right)^{a+b-1}\alpha b F \frac{\p}{\p\eta}\left(\frac{1}{\theta_f}\frac{\p F}{\p\eta}\right) = \frac{\p}{\p\eta}\left(\frac{\mu}{T}\frac{\p^2F}{\p\eta^2}\right) + \\
    - \left(\alpha x\right)^{3b-a} \left(\frac{d_{g0}^\ast}{d_g^\ast}\right)^2 \left[ C_D\frac{\left(1-\theta_f\right)^{2}}{\theta_f^2} \mu T \frac{\p F}{\p\eta} + C_F \frac{1-\theta_f}{\theta_f^2}\left(\frac{\p F}{\p\eta}\right)^2\right] .
\end{multline}
where the symbol $\left.\p/\p x\right\vert_{\eta}$ denotes the derivative with respect to $x$ at constant $\eta$. The self-similar solution \eqref{eq:self-similar_porous_momentum} is recovered if $a=b=1/2$, $\alpha=2$ and $d_g^\ast/d_{g0}^\ast = \left(2x\right)^{1/2}$ \cite{Tsiberkin_2018a}. 

%%%%%%%%%%%%%%%%%%%%%%%%%%%%%%%%%%%%%%%%%%%%%%%%%%%%
%%%%%%%%%%%%%%%%%%%%%%%%%%%%%%%%%%%%%%%%%%%%%%%%%%%%
%%%%%%%%%%%%%%%%%%%%%%%%%%%%%%%%%%%%%%%%%%%%%%%%%%%%
%%%%%%%%%%%%%%%%%%%%%%%%%%%%%%%%%%%%%%%%%%%%%%%%%%%%
%%%%%%%%%%%%%%%%%%%%%%%%%%%%%%%%%%%%%%%%%%%%%%%%%%%%

\section{Numerical procedures}\label{app:numerical_method}
\begin{sloppypar}
The governing equations are solved using the block-elimination method described in \cite{Keller_Cebeci_1971,Keller_Cebeci_1972} and \cite{Cebeci_2002}. The ordinary differential equations \eqref{eq:self-similar_porous} are decomposed into a system of first order ODEs
\begin{subequations}\label{eq:self-similar_numerics}
\begin{align}
    \theta_f^2\left(bv\right)^\prime - \zeta_ffu + \theta_ffv - \left(1-2\theta_f+\theta_f^2\right)ou - \left(1-\theta_f\right)qu^2 & = 0, \\
    \theta_f\left(\theta_{\p ff}ep\right)^\prime + \theta_f fp + dv^2 & = 0,
\end{align}
\end{subequations}
with the auxiliary equations $u=f^\prime$, $v=u^\prime$ and $p=g^\prime$. Here,
\begin{subequations}
\begin{equation}\label{eq:numerics_variables}
\begin{array}{ccccc}
    f = F, & u = F^\prime, & v=F^{\prime\prime}, & g= T, & p=T^\prime, \\
\end{array}
\end{equation}
\begin{equation}
\begin{array}{ccccc}
    \displaystyle b=\frac{\mu}{g}, & \displaystyle e= \frac{1}{\Prandtl}\frac{\mu}{g}, & \displaystyle d = \left(\gamma-1\right)\Mach^2\frac{\mu}{g},& \displaystyle o = C_Dbg^2, & \displaystyle q= C_F,
\end{array}
\end{equation}
\end{subequations}
and $\zeta_f=\theta_f^\prime$. The nonlinearity of the first-order system \eqref{eq:self-similar_numerics} is treated by Taylor-expanding the variables \eqref{eq:numerics_variables} into the residuals
\begin{equation}\label{eq:residuals}
\begin{array}{ccccc}
    \delta f = f - f^{(0)}, & \delta u = u - u^{(0)}, & \delta v = v - v^{(0)}, & \delta g = g - g^{(0)}, & \delta p = p - p^{(0)},
\end{array}
\end{equation}
into the equations. The momentum and enthalpy balances \eqref{eq:self-similar_numerics} and the auxiliary equations reduce to linearized equations for the residuals. The domain is discretized using uniform second-order finite differences centered in the midpoint $j-1/2$, where $0\leq j\leq N$ and $N$ is the number of grid points along $\eta$
\begin{subequations}\label{eq:self-similar_numerics_block-elimination}
\begin{equation}
     \delta f_j - \delta f_{j-1} - \frac{h}{2}\left(\delta u_j + \delta u_{j-1}\right) = f_{j-1}^{(0)} - f_j^{(0)} + \frac{h}{2}\left(u_j^{(0)} + u_{j-1}^{(0)}\right) \equiv \left(r_1^{(0)}\right)_j ,
\end{equation}
\begin{equation}
     \left(s_1\right)_j \delta v_j + \left(s_2\right)_j \delta v_{j-1} + \left(s_3\right)_j \delta f_j + \left(s_4\right)_j \delta f_{j-1} + \left(s_5\right)_j \delta u_j + \left(s_6\right)_j \delta u_{j-1} = \left(r_2^{(0)}\right)_j  ,\label{eq:numerical_momentum}
\end{equation}
\begin{multline}
     \left(\beta_1\right)_j \delta p_j + \left(\beta_2\right)_j \delta p_{j-1} + \left(\beta_3\right)_j \delta f_j + \left(\beta_4\right)_j \delta f_{j-1} + \left(\beta_5\right)_j \delta u_j + \\
     + \left(\beta_6\right)_j \delta u_{j-1} + \left(\beta_7\right)_j \delta g_j + \left(\beta_8\right)_j \delta g_{j-1} + \left(\beta_9\right)_j \delta v_j + \left(\beta_{10}\right)_j \delta v_{j-1} = \left(r_3^{(0)}\right)_j , \label{eq:numerical_enthalpy}
\end{multline} 
\begin{equation}
     \delta u_{j+1} - \delta u_j - \frac{h}{2}\left(\delta u_{j+1} + \delta u_j\right) = u_j^{(0)} - u_{j+1}^{(0)} + \frac{h}{2}\left(u_{j+1}^{(0)} + u_j^{(0)}\right) \equiv \left(r_4^{(0)}\right)_j ,
\end{equation}
\begin{equation}
     \delta g_{j+1} - \delta g_j - \frac{h}{2}\left(\delta p_{j+1} + \delta p_j\right) = g_j^{(0)} - g_{j+1}^{(0)} + \frac{h}{2}\left(p_{j+1}^{(0)} + p_j^{(0)}\right) \equiv \left(r_5^{(0)}\right)_j ,
\end{equation}
\end{subequations}
where $h=\eta_j-\eta_{j-1}$ is the grid size. The coefficients in \eqref{eq:numerical_momentum} are
\begin{subequations}
\begin{equation}
     \left(s_1\right)_j = \left(\theta_f^2\right)_{j-1/2} h^{-1}b_j + \left(\theta_f\right)_{j-1/2}\frac{f_j^{(0)}}{2}, 
\end{equation}
\begin{equation}
     \left(s_2\right)_j = - \left(\theta_f^2\right)_{j-1/2} h^{-1}b_{j-1} + \left(\theta_f\right)_{j-1/2}\frac{f_{j-1}^{(0)}}{2}, 
\end{equation}
\begin{equation}
     \left(s_3\right)_j = -\left(\zeta_f\right)_{j-1/2}\frac{u_j^{(0)}}{2} +\left(\theta_f\right)_{j-1/2}\frac{v_j^{(0)}}{2}, 
\end{equation}
\begin{equation}
     \left(s_4\right)_j = -\left(\zeta_f\right)_{j-1/2}\frac{u_{j-1}^{(0)}}{2} +\left(\theta_f\right)_{j-1/2}\frac{v_{j-1}^{(0)}}{2}, 
\end{equation}
\begin{multline}
     \left(s_5\right)_j = -\left(\zeta_f\right)_{j-1/2}\frac{f_j^{(0)}}{2} - \left[1-2\left(\theta_f\right)_{j-1/2}+\left(\theta_f^2\right)_{j-1/2}\right]\frac{o_j}{2} + \\
     - \left[1-\left(\theta_f\right)_{j-1/2}\right]\left(qu^{(0)}\right)_j, 
\end{multline}
\begin{multline}
     \left(s_6\right)_j = -\left(\zeta_f\right)_{j-1/2}\frac{f_{j-1}^{(0)}}{2} - \left[1-2\left(\theta_f\right)_{j-1/2}+\left(\theta_f^2\right)_{j-1/2}\right]\frac{o_{j-1}}{2} +\\
     - \left[1-\left(\theta_f\right)_{j-1/2}\right]\left(qu^{(0)}\right)_{j-1},
\end{multline}
\begin{multline}
     \left(r_2\right)_j = -\left(\theta_f^2\right)_{j-1/2}h^{-1}\left[\left(bv^{(0)}\right)_j - \left(bv^{(0)}\right)_{j-1}\right] + \left(\zeta_f\right)_{j-1/2}\left(f^{(0)}u^{(0)}\right)_{j-1/2} + \\
     - \left(\theta_f\right)_{j-1/2}\left(f^{(0)}v^{(0)}\right)_{j-1/2} + \left[1-2\left(\theta_f\right)_{j-1/2}+\left(\theta_f^2\right)_{j-1/2}\right]\left(ou^{(0)}\right)_{j-1/2} + \\
     + \left[1-\left(\theta_f\right)_{j-1/2}\right]\left(qu^{(0)2}\right)_{j-1/2},
\end{multline}
\end{subequations}
and the coefficients in \eqref{eq:numerical_enthalpy} are
\begin{subequations}
\begin{equation}
    \left(\beta_1\right)_j = \left(\theta_f\right)_{j-1/2} \left[h^{-1}\left(\theta_{\p ff}e\right)_j + \frac{f_j^{(0)}}{2}\right], 
\end{equation}
\begin{equation}
    \left(\beta_2\right)_j = \left(\theta_f\right)_{j-1/2} \left[-h^{-1}\left(\theta_{\p ff}e\right)_{j-1} + \frac{f_{j-1}^{(0)}}{2}\right], 
\end{equation}
\begin{equation}
    \left(\beta_3\right)_j = \left(\theta_f\right)_{j-1/2}\frac{p^{(0)}_j}{2}, 
\end{equation}
\begin{equation}
    \left(\beta_4\right)_j = \left(\theta_f\right)_{j-1/2}\frac{p^{(0)}_{j-1}}{2}, 
\end{equation}
\begin{equation}
    \left(\beta_5\right)_j = \left(\beta_6\right)_j = \left(\beta_7\right)_j = \left(\beta_8\right)_j = 0,
\end{equation}
\begin{equation}
    \left(\beta_9\right)_j = \left(dv^{(0)}\right)_j, 
\end{equation}
\begin{equation}
    \left(\beta_{10}\right)_j = \left(dv^{(0)}\right)_{j-1}, 
\end{equation}
\begin{multline}
    \left(r_3\right)_j = -\left(\theta_f\right)_{j-1/2}h^{-1}\left[\left(\theta_{\p ff}ep^{(0)}\right)_j - \left(\theta_{\p ff}ep^{(0)}\right)_{j-1}\right] +\\
    - \left(\theta_f\right)_{j-1/2}\left(f^{(0)}p^{(0)}\right)_{j-1/2} - \left(dv^{(0)2}\right)_{j-1/2}.
\end{multline}
\end{subequations}
The system \eqref{eq:self-similar_numerics_block-elimination} is elliptic in $\eta$. Three boundary conditions $\delta f_0=\delta u_0=\delta p_0=0$ are imposed at the bottom solid wall $j=0$ and two boundary conditions $\delta u_{N-1}=\delta g_{N-1}=0$ are imposed in the free stream $j=N-1$. Following \cite{Cebeci_2002}, the system \eqref{eq:self-similar_numerics_block-elimination} is written in the matrix form 
\begin{equation}\label{eq:thomas_algorithm}
     \boldsymbol{\underline{\underline{B}}}_{j} \boldsymbol{\underline{\delta}}_{j-1} + \boldsymbol{\underline{\underline{A}}}_{j} \boldsymbol{\underline{\delta}}_{j} + \boldsymbol{\underline{\underline{C}}}_{j} \boldsymbol{\underline{\delta}}_{j+1} = \boldsymbol{\underline{r}^{(0)}}_j,
\end{equation}
where $\boldsymbol{\underline{\delta}} = [\begin{array}{ccccc}
    \delta f & \delta u & \delta v & \delta g & \delta p\end{array}]^T$ is the vector of the residuals, $\boldsymbol{\underline{r}^{(0)}} = [\begin{array}{ccccc}
    r_1^{(0)} & r_2^{(0)} & r_3^{(0)} & r_4^{(0)} & r_5^{(0)}\end{array}]^T$ is the vector of the equations for the variables $f^{(0)}$, $u^{(0)}$, $v^{(0)}$, $g^{(0)}$ and $p^{(0)}$, and $\boldsymbol{\underline{\underline{B}}}_{j}$, $\boldsymbol{\underline{\underline{A}}}_{j}$ and $\boldsymbol{\underline{\underline{C}}}_{j}$ are the coefficient matrices. A tridiagonal block-elimination (Thomas) algorithm algorithm is employed to compute the vector of the residuals $\boldsymbol{\underline{\delta}}$ in \eqref{eq:thomas_algorithm}. The residuals are then used to update $f^{(0)}$, $u^{(0)}$, $v^{(0)}$, $g^{(0)}$ and $p^{(0)}$ through \eqref{eq:residuals} and the whole procedure is repeated iteratively until the magnitude of $\abs{\boldsymbol{\underline{\delta}}}$ falls below the prescribed tolerance $10^{-12}$. 
\end{sloppypar}

%%%%%%%%%%%%%%%%%%%%%%%%%%%%%%%%%%%%%%%%%%%%%%%%%%%%%%%%%%%%%%%%%%%%
%%%%%%%%%%%%%%%%%%%%%%%%%%%%%%%%%%%%%%%%%%%%%%%%%%%%%%%%%%%%%%%%%%%%
%%%%%%%%%%%%%%%%%%%%%%%%%%%%%%%%%%%%%%%%%%%%%%%%%%%%%%%%%%%%%%%%%%%%
%%%%%%%%%%%%%%%%%%%%%%%%%%%%%%%%%%%%%%%%%%%%%%%%%%%%%%%%%%%%%%%%%%%%
%%%%%%%%%%%%%%%%%%%%%%%%%%%%%%%%%%%%%%%%%%%%%%%%%%%%%%%%%%%%%%%%%%%%

\printbibliography

@book{Anderson_2019,
  title={Hypersonic and high-temperature gas dynamics},
  author={Anderson, J. D.},
  isbn={9781624105142},
  lccn={2018061021},
  year={2019},
  publisher={AIAA, Inc.}
}

@article{Bachmat_Bear_1986,
  title={Macroscopic modelling of transport phenomena in porous media. 1: the continuum approach},
  author={Bachmat, Y. and Bear, J.},
  journal={Transp. Porous Media},
  volume={1},
  number={3},
  pages={213--240},
  year={1986},
  publisher={Springer},
  issn={1573-1634},
  doi={10.1007/BF00238181}
}

@article{Barrere_Gipouloux_Whitaker_1992,
  title = {On the closure problem for {Darcy's} law},
  author = {Barr\`{e}re, J. and Gipouloux, O. and Whitaker, S.},
  year = {1992},
  month = {03},
  journal = {Transp. Porous Media},
  volume = {7},
  number = {3},
  pages = {209--222},
  issn = {1573-1634},
  doi = {10.1007/BF01063960},
}

@book{Bear_Bachmat_1990,
  title={Introduction to modeling of transport phenomena in porous media},
  author={Bear, J. and Bachmat, Y.},
  year={1990},
  isbn={9789400919266},
  publisher = {Kluwer Acad. Publ.},
  address   = {Dordrecht}
}

@phdthesis{Breugem_2005,
  author  = {Breugem, W. P.},
  title   = {The influence of wall permeability on laminar and turbulent flows: theory and simulations},
  school  = {TU Delft},
  year    = {2005}
}

@Article{Breugem_Boersma_Uittenbogaard_2005,
author={Breugem, W. P.
and Boersma, B. J.
and Uittenbogaard, R. E.},
title={The laminar boundary layer over a permeable wall},
journal={Transp. Porous Media},
year={2005},
month={06},
day={01},
volume={59},
number={3},
pages={267-300},
issn={1573-1634},
doi={10.1007/s11242-004-2557-1}
}

@article{Breugem_Boersma_2005,
author = {Breugem, W. P.  and Boersma, B. J. },
title = {Direct numerical simulations of turbulent flow over a permeable wall using a direct and a continuum approach},
journal = {Phys. Fluids},
volume = {17},
number = {2},
pages = {025103},
year = {2005},
doi={10.1063/1.1835771}
}

@article{Breugem_Boersma_Uittenbogaard_2006, 
title={The influence of wall permeability on turbulent channel flow}, 
volume={562},
journal={J. Fluid Mech.}, 
publisher={Cambridge University Press}, 
author={Breugem, W. P. and Boersma, B. J. and Uittenbogaard, R. E.}, 
year={2006}, 
pages={35–72},
doi={10.1017/S0022112006000887}
}

@book{Cebeci_2002,
  title={Convective heat transfer},
  author={Cebeci, T.},
  year={2002},
  publisher={Horizons Publ.},
  address={Heidelberg}
}

@article{Celli_Rees_Barletta_2010,
title = {The effect of local thermal non-equilibrium on forced convection boundary layer flow from a heated surface in porous media},
journal = {Int. J. Heat Mass Transfer},
volume = {53},
number = {17},
pages = {3533-3539},
year = {2010},
issn = {0017-9310},
doi = {10.1016/j.ijheatmasstransfer.2010.04.014},
author = {Celli, M. and Rees, D.A.S. and Barletta, A.},
}

@article{Costa_2006,
author = {Costa, A.},
title = {Permeability-porosity relationship: a reexamination of the Kozeny-Carman equation based on a fractal pore-space geometry assumption},
journal = {Geophys. Res. Lett.},
volume = {33},
number = {2},
year = {2006}
}

@article{Emanuel_Jones_1968,
title = {Compressible flow through a porous plate},
journal = {Int. J. Heat Mass Transf.},
volume = {11},
number = {5},
pages = {827-836},
year = {1968},
issn = {0017-9310},
doi = {10.1016/0017-9310(68)90127-0},
author = {Emanuel, G. and Jones, J. P.}
}

@article{Goharzadeh_Khalili_Jorgensen_2005,
    author = {Goharzadeh, A. and Khalili, A. and Jørgensen, B. B.},
    title = {Transition layer thickness at a fluid-porous interface},
    journal = {Phys. Fluids},
    volume = {17},
    number = {5},
    pages = {057102},
    year = {2005},
    month = {04},
    issn = {1070-6631},
    doi = {10.1063/1.1894796}
}

@article{Gray_1975,
title = {A derivation of the equations for multi-phase transport},
journal = {Chem. Eng. Sci.},
volume = {30},
number = {2},
pages = {229-233},
year = {1975},
issn = {0009-2509},
doi = {10.1016/0009-2509(75)80010-8},
author = {Gray, W. G.}
}

@article{Graziosi_Brown_2002, 
title={Experiments on stability and transition at {Mach} 3}, 
volume={472}, 
journal={J. Fluid Mech.}, 
publisher={Cambridge University Press}, 
author={Graziosi, P. and Brown, G. L.}, 
year={2002}, 
pages={83–124},
doi={10.1017/S0022112002002094}
}

@inbook{Innocentini_Sepulveda_Ortega_2006,
  editor    = {Scheffler, M. and Colombo, P.},
  author    = {Innocentini, M. D. M. and Sepulveda, P. and Ortega, F. S.},
  booktitle = {Cellular ceramics: structure, manufacturing, properties and applications},
  chapter   = {4.2},
  title     = {Permeability},
  publisher = {Wiley-VCH},
  year      = {2006},
  isbn      = {9783527313204}
}

@InProceedings{Keller_Cebeci_1971,
author={Keller, H. B.
and Cebeci, T.},
editor={Holt, M.},
title={Accurate numerical methods for boundary layer flows I: two dimensional laminar flows},
booktitle={Proc. Second Int. Conf. Numer. Methods Fluid Dyn.},
year={1971},
publisher={Springer Berlin Heidelberg},
pages={92--100}
}

@article{Keller_Cebeci_1972,
author = {Keller, H. B. and Cebeci, T.},
title = {Accurate numerical methods for boundary-layer flows II: two dimensional turbulent flows},
journal = {AIAA J.},
volume = {10},
number = {9},
pages = {1193-1199},
year = {1972}
}

@article{Harter_Martinez_Poser_Weigand_Lamanna_2023,
    author = {H\"{a}rter, J. and Mart\'{i}nez, D. S. and Poser, R. and Weigand, B. and Lamanna, G.},
    title = "{Coupling between a turbulent outer flow and an adjacent porous medium: high resolved particle image velocimetry measurements}",
    journal = {Phys. Fluids},
    volume = {35},
    number = {2},
    year = {2023},
    month = {02},
    issn = {1070-6631},
    doi={10.1063/5.0132193}
}

@article{Kaviany_1987,
    author = {Kaviany, M.},
    title = {Boundary-layer treatment of forced convection heat transfer from a semi-infinite flat plate embedded in porous media},
    journal = {J. Heat Transf.},
    volume = {109},
    number = {2},
    pages = {345-349},
    year = {1987},
    month = {05},
    issn = {0022-1481},
    doi = {10.1115/1.3248086}
}

@article{Khalifa_Pocher_Tilton_2020,
title = {Regimes of flow through cylinder arrays subject to steady pressure gradients},
journal = {Int. J. Heat Mass Transf.},
volume = {159},
pages = {120072},
year = {2020},
issn = {0017-9310},
author = {Khalifa, Z. and Pocher, L. and Tilton, N.},
doi = {j.ijheatmasstransfer.2020.120072}
}

@article{Lasseux_Valdes-Parada_2017,
title = {On the developments of {Darcy}'s law to include inertial and slip effects},
journal = {C. R. M\'{e}c.},
volume = {345},
number = {9},
pages = {660-669},
year = {2017},
issn = {1631-0721},
author = {Lasseux, D. and Vald\'{e}s-Parada, F. J.},
doi = {10.1016/j.crme.2017.06.005}
}

@article{Maslov_Shiplyuk_Sidorenko_Arnal_2001, 
title={Leading-edge receptivity of a hypersonic boundary layer on a flat plate}, 
volume={426}, 
journal={J. Fluid Mech.}, 
publisher={Cambridge University Press}, 
author={Maslov, A. A. and Shiplyuk, A. N. and Sidorenko, A. A. and Arnal, D.}, 
year={2001}, 
pages={73–94},
doi={10.1017/S0022112000002147}
}

@article{Maslov_Mironov_Poplavskaya_Kirilovskiy_2019, 
title={Supersonic flow around a cylinder with a permeable high-porosity insert: experiment and numerical simulation}, 
volume={867}, 
DOI={10.1017/jfm.2019.165}, 
journal={J. Fluid Mech.}, 
publisher={Cambridge University Press}, 
author={Maslov, A. A. and Mironov, S. G. and Poplavskaya, T. V. and Kirilovskiy, S. V.}, 
year={2019}, 
pages={611–632}
}

@article{Mironov_Maslov_Poplavskaya_Kirilovskiy_2015,
  title = {Modeling of a supersonic flow Around a cylinder with a gas-permeable porous insert},
  author = {Mironov, S. G. and Maslov, A. A. and Poplavskaya, T. V. and Kirilovskiy, S. V.},
  year = {2015},
  month = {07},
  journal = {J. Appl. Mech. Tech. Phys.},
  volume = {56},
  number = {4},
  pages = {549--557},
  issn = {1573-8620},
  doi = {10.1134/S0021894415040021},
}

@article{Nakayama_Kokudai_Koyama_1990,
    author = {Nakayama, A. and Kokudai, T. and Koyama, H.},
    title = {Non-{Darcian} boundary layer flow and forced convective heat transfer over a flat plate in a fluid-saturated porous medium},
    journal = {J. Heat Transf.},
    volume = {112},
    number = {1},
    pages = {157-162},
    year = {1990},
    month = {02},
    issn = {0022-1481},
    doi={10.1115/1.2910338}
}

@article{Neale_Nader_1974,
author = {Neale, G. and Nader, W.},
title = {Practical significance of {Brinkman's} extension of {Darcy's} law: coupled parallel flows within a channel and a bounding porous medium},
journal = {Can. J. Chem. Eng.},
volume = {52},
number = {4},
pages = {475-478},
doi = {10.1002/cjce.5450520407},
year = {1974}
}

@article{Negi_Mishra_Skote_2015,
  title    = {{DNS} of a single {low-speed} streak subject to spanwise wall oscillations},
  author   = {Negi, P. S. and Mishra, M. and Skote, M.},
  journal  = {Flow Turbul. Combust.},
  volume   =  {94},
  number   =  {4},
  pages    = {795--816},
  month    =  {06},
  year     =  {2015},
  doi      = {10.1007/s10494-015-9599-z}
}

@Article{Nield_1994,
author={Nield, D. A.},
title={Modelling high speed flow of a compressible fluid in a saturated porous medium},
journal={Transp. Porous Media},
year={1994},
month={01},
day={01},
volume={14},
number={1},
pages={85-88},
issn={1573-1634},
doi = {10.1007/BF00617029}
}

@book{Nield_Bejan_2017,
  title={Convection in porous media},
  author={Nield, D. A. and Bejan, A.},
  isbn={9783319495620},
  year={2017},
  publisher={Springer Int. Publ.},
  address={New York}
}

@article{Nield_1991,
    title = {The limitations of the {Brinkman-Forchheimer} equation in modeling flow in a saturated porous medium and at an interface},
    journal = {Int. J. Heat Fluid Flow},
    volume = {12},
    number = {3},
    pages = {269-272},
    year = {1991},
    issn = {0142-727X},
    doi = {10.1016/0142-727X(91)90062-Z},
    author = {Nield, D. A.},
}

@article{Nield_Kutsenov_2003,
  title = {Boundary-layer analysis of forced convection with a plate and porous substrate},
  author = {Nield, D. A. and Kuznetsov, A. V.},
  year = {2003},
  month = {12},
  journal = {Acta Mech.},
  volume = {166},
  number = {1},
  pages = {141--148},
  issn = {1619-6937},
  doi = {10.1007/s00707-003-0050-5}
}

@article{Ochoa-Tapia_Whitaker_1995a,
title = {Momentum transfer at the boundary between a porous medium and a homogeneous fluid—{I}. {Theoretical} development},
journal = {Int. J. Heat Mass Transf.},
volume = {38},
number = {14},
pages = {2635-2646},
year = {1995},
issn = {0017-9310},
doi = {10.1016/0017-9310(94)00346-W},
author = {Ochoa-Tapia, J. A. and Whitaker, S.}
}

@article{Papalexandris_2023a,
  title = {Boundary-layer flow in a porous domain above a flat plate},
  author = {Papalexandris, M. V.},
  year = {2023},
  month = {05},
  journal = {J. Eng. Math.},
  volume = {140},
  number = {1},
  pages = {4},
  issn = {1573-2703},
  doi = {10.1007/s10665-023-10269-4}
}

@Article{Papalexandris_2023b,
author={Papalexandris, M. V.},
title={Thermal boundary-layer solutions for forced convection in a porous domain above a flat plate},
journal={J. Eng. Math.},
year={2023},
month={12},
day={08},
volume={144},
number={1},
pages={3},
issn={1573-2703},
doi={10.1007/s10665-023-10311-5}
}

@Article{Quintard_Whitaker_1994,
author={Quintard, M.
and Whitaker, S.},
title={Transport in ordered and disordered porous media {II}: generalized volume averaging},
journal={Transp. Porous Media},
year={1994},
month={02},
day={01},
volume={14},
number={2},
pages={179-206},
doi={10.1007/BF00615200}
}

@article{Running_Bemis_Hill_Borg_Redmond_Jantze_Scalo_2023,
  title = {Attenuation of hypersonic second-mode boundary-layer instability with an ultrasonically absorptive silicon-carbide foam},
  author = {Running, C. L. and Bemis, B. L. and Hill, J. L. and Borg, M. P. and Redmond, J. J. and Jantze, K. and Scalo, C.},
  year = {2023},
  month = {03},
  journal = {Exp. Fluids},
  volume = {64},
  number = {4},
  pages = {79},
  issn = {1432-1114},
  doi = {10.1007/s00348-023-03615-w}
}

@article{Shreeve_1968,
author = {Shreeve, R. P.},
title = {Supersonic flow from a porous metal plate.},
journal = {AIAA J.},
volume = {6},
number = {4},
pages = {752-753},
year = {1968},
doi = {10.2514/3.4589}
}

@article{Sparrow_Quack_Boerner_1970,
author = {Sparrow, E. M. and Quack, H. and Boerner, C. J.},
title = {Local nonsimilarity boundary-layer solutions},
journal = {AIAA J.},
volume = {8},
number = {11},
pages = {1936-1942},
year = {1970},
doi = {10.2514/3.6029}
}

@book{Stewartson_1964,
  title={The theory of laminar boundary layers in compressible fluids},
  author={Stewartson, K.},
  year={1964},
  publisher={Clarendon Press},
  address = {Oxford}
}

@article{Sorek_Levi-Hevroni_Levy_Ben-Dor_2005,
  title = {Extensions to the macroscopic {Navier–Stokes} equation},
  author = {Sorek, S. and Levi-Hevroni, D. and Levy, A. and Ben-Dor, G.},
  year = {2005},
  month = {11},
  day = {01},
  journal = {Transp. Porous Media},
  volume = {61},
  number = {2},
  pages = {215--233},
  issn = {1573-1634},
  doi={10.1007/s11242-004-7906-6}
}

@article{Tilton_Cortelezzi_2008, 
title={Linear stability analysis of pressure-driven flows in channels with porous walls}, 
volume={604}, 
DOI={10.1017/S0022112008001341}, 
journal={J. Fluid Mech.}, 
publisher={Camb. Univ. Press}, 
author={Tilton, N. and Cortelezzi, L.}, 
year={2008}, 
pages={411–445}
}

@article{Tilton_Cortelezzi_2015,
author = {Tilton, N. and Cortelezzi, L.},
title = {Stability of boundary layers over porous walls with suction},
journal = {AIAA J.},
volume = {53},
number = {10},
pages = {2856-2868},
year = {2015},
doi = {10.2514/1.J053716}
}

@Article{Tsiberkin_2016,
author={Tsiberkin, K.},
title={On the structure of the steady-state flow velocity field near the interface between a homogeneous liquid and a {Brinkman} porous medium},
journal={Tech. Phys.},
year={2016},
month={08},
day={01},
volume={61},
number={8},
pages={1181-1186},
issn={1090-6525},
doi={10.1134/S1063784216080272}
}

@article{Tsiberkin_2018a,
  title={Effect of inertial terms on fluid–porous medium flow coupling},
  author={Tsiberkin, K.},
  journal={Transp. Porous Media},
  volume={121},
  number={1},
  pages={109--120},
  year={2018},
  publisher={Springer},
  issn={1573-1634},
  doi={10.1007/s11242-017-0951-8}
}

@article{Tsiberkin_2018b,
  title={Inertial and {Darcy}’s terms ratio in boundary layer at fluid–porous medium interface},
  author={Tsiberkin, K.},
  journal={Transp. Porous Media},
  volume={125},
  number={2},
  pages={259--269},
  year={2018},
  publisher={Springer},
  issn={1573-1634},
  doi={10.1007/s11242-018-1117-z}
}

@article{de-Ville_1996,
  title = {On the properties of compressible gas flow in a porous media},
  author = {{\DE{Ville}{de}{de}} Ville, A.},
  year = {1996},
  month = {03},
  journal = {Transp. Porous Media},
  volume = {22},
  number = {3},
  pages = {287--306},
  issn = {1573-1634},
  doi={10.1007/BF00161628}
}

@article{Vafai_Kim_1990,
    author = {Vafai, K. and Kim, S.-J.},
    title = "{Analysis of surface enhancement by a porous substrate}",
    journal = {J. Heat Transf.},
    volume = {112},
    number = {3},
    pages = {700-706},
    year = {1990},
    month = {08},
    issn = {0022-1481},
    doi = {10.1115/1.2910443}
}

@book{van-Dyke_1975,
  title={Perturbation methods in fluid mechanics},
  author={{\VAN{Dyke}{van}{van}} Dyke, M.},
  lccn={76351560},
  year={1975},
  publisher={Parabolic Press},
  address = {Stanford}
}

@article{Wedin_Cherubini_2016,
    doi = {10.1088/0169-5983/48/6/061411},
    year = {2016},
    month = {11},
    publisher = {IOP Publ.},
    volume = {48},
    number = {6},
    pages = {061411},
    author = {Wedin, H. and Cherubini, S.},
    title = {Permeability models affecting nonlinear stability in the asymptotic suction boundary layer: the {Forchheimer} versus the {Darcy} model},
    journal = {Fluid Dyn. Res.}
}

@article{Whitaker_1969,
  title={Advances in theory of fluid motion in porous media},
  author={Whitaker, S.},
  journal={Ind. Eng. Chem.},
  volume={61},
  number={12},
  pages={14--28},
  year={1969},
  publisher={American Chemical Society},
  issn={0019-7866},
  doi={10.1021/ie50720a004}
}

@article{Whitaker_1986,
  title = {Flow in porous media {I}: a theoretical derivation of {Darcy}'s law},
  author = {Whitaker, S.},
  year = {1986},
  month = {03},
  day = {01},
  journal = {Transp. Porous Media},
  volume = {1},
  number = {1},
  pages = {3--25},
  issn = {1573-1634},
  doi = {10.1007/BF01036523}
}

@article{Whitaker_1996,
  title = {The {Forchheimer} equation: a theoretical development},
  author = {Whitaker, S.},
  year = {1996},
  month = {10},
  day = {01},
  journal = {Transp. Porous Media},
  volume = {25},
  number = {1},
  pages = {27--61},
  issn = {1573-1634},
  doi= {10.1007/BF00141261}
}

@book{Whitaker_1998,
  title={The method of volume averaging},
  author={Whitaker, S.},
  isbn={9780792354864},
  lccn={98047438},
  year={1998},
  publisher={Kluwer Acad. Publ.},
  address   = {Dordrecht}
}

@article{Wu_Mirbod_2018,
author = {Wu, Z.  and Mirbod, P. },
title = {Experimental analysis of the flow near the boundary of random porous media},
journal = {Phys. Fluids},
volume = {30},
number = {4},
pages = {047103},
year = {2018},
doi={10.1063/1.5021903}
}

\end{document}